%% file: kdiff_main.tex
\documentclass[journal]{IEEEtran}
\usepackage{relsize}
\usepackage{amsmath,amssymb,amsthm}
\usepackage{amsfonts}
\usepackage{graphicx}
\usepackage{cite}
\usepackage{pgf}
\usepackage{tikz}
\usepackage{rotating}
\usepackage{setspace}
\usepackage{enumerate}
\usepackage{pifont}
\usepackage{mathtools}
\usetikzlibrary{arrows,automata,patterns,decorations.pathreplacing,positioning,calc}
\usepackage{algorithm}
\usepackage{algorithmic}
\usepackage{bm}
\usepackage{color}
\newcommand{\blue}[1]{#1}
\newcommand{\red}[1]{#1}
\newcommand{\med}[1]{#1}
\newcommand{\pmed}[1]{#1}
\newcommand{\picsize}{0.6\columnwidth}

\newtheorem{approximation}{Approximation}
\newtheorem{property}{Property}

\newtheoremstyle{ex}
  {5pt} 
  {5pt} 
  {\small} 
  {10pt} 
  {\itshape} 
  {:} 
  {5pt plus 1pt minus 1pt} 
  {} 
\theoremstyle{ex}
\newtheorem*{example}{Example}

\newtheoremstyle{def}
  {\topsep} 
  {\topsep} 
  {} 
  {0pt} 
  {\scshape} 
  {:} 
  {5pt plus 1pt minus 1pt} 
  {} 
\theoremstyle{def}
\newtheorem{definition}{Definition}

\newcommand{\beq}{\begin{equation}}
\newcommand{\eeq}{\end{equation}}
\newcommand{\beqa}{\begin{eqnarray}}
\newcommand{\eeqa}{\end{eqnarray}}
\newcommand{\pkt}{\textnormal{\textsf{p}}} 

\newcommand{\tablespace}{\vspace{2mm}}

\newcommand{\rx}{\mathrm{Rx}}
\newcommand{\tick}{\mbox{\ding{52}}}
\newcommand{\cross}{\mbox{\ding{56}}}
\newcommand{\nnt}[1]{N_{#1}(t)} 
\newcommand{\nn}[2]{N_{#1}(#2)}
\newcommand{\buf}{\textnormal{\textbf{b}}}
\newcommand{\ith}{^{\mbox{\smaller{th}}}}
\newcommand{\mubar}{\overline{\mu}}
\newcommand{\deliv}{d}
\newcommand{\bufpkt}{\textnormal{\textbf{B}}}
\newcommand{\betav}{\boldsymbol{\beta}}
\newcommand{\muv}{\boldsymbol{\mu}}
\newcommand{\scaling}{\omega}

\newcommand{\boxwidth}{0.4cm}
\newcommand{\boxheight}{.9cm}
\newcommand{\boxrxedcol}{black!90}
\newcommand{\drawrxbox}[2]{
    \foreach \x in {{#1},...,{#2}}
      {\draw (\boxwidth*\x,0cm) node[rectangle,draw,minimum width=\boxwidth,minimum height=\boxheight,fill=\boxrxedcol]{};}
 } 
 \newcommand{\drawmtbox}[2]{
    \foreach \x in {{#1},...,{#2}}
      {\draw (\boxwidth*\x,0cm) node[rectangle,draw,minimum width=\boxwidth,minimum height=\boxheight]{};}
 } 
\newcommand{\drawpatternbox}[3]{
    \foreach \x in {{#1},...,{#2}}
      {\draw (\boxwidth*\x,0cm) node[rectangle,draw,minimum width=\boxwidth,minimum height=\boxheight,fill={#3}]{};}
  \foreach \x in {{#1},...,{#2}}
      {\draw (\boxwidth*\x,0cm) node[rectangle,draw,minimum width=\boxwidth,minimum height=\boxheight,pattern=north west lines, pattern color=black!10]{};}
}

\newcommand{\rectheight}{0.7cm}
\newcommand{\rectwidth}{0.9cm}
\newcommand{\drawrect}[5]{
  \definecolor{tempcolor}{rgb}{#5,#5,#5}
  \draw [fill=tempcolor] (#2*\rectwidth,-#1*\rectheight) rectangle (#3*\rectwidth+#2*\rectwidth,-#1*\rectheight-\rectheight) node[midway, anchor=center] {#4};
}
\newcommand{\drawnbrect}[5]{
  \definecolor{tempcolor}{rgb}{#5,#5,#5}
  \draw [fill=tempcolor] (#2*\rectwidth,-#1*\rectheight) rectangle (#3*\rectwidth+#2*\rectwidth,-#1*\rectheight-\rectheight) node[midway, anchor=center,black] {#4};
}

\newcommand{\drawlabelbox}[2]{
 \draw ({#1},0) node[rectangle,draw,minimum width=\boxwidth,minimum height=\boxheight,label=above:{#2}]{};}
\newcommand{\belowbrace}[3]{
\draw [thick,decoration= {brace,mirror,amplitude=0.2cm,raise=0.65cm}, decorate] ({#1},0cm)--({#2},0cm)
node[midway,below,yshift=-0.8cm,minimum width=1.8cm,align=center]{#3};}

\begin{document}
\onecolumn
\title{Beyond the Min-Cut Bound: Deterministic Network Coding for Asynchronous Multirate Broadcast}
\author{\IEEEauthorblockN{Amy Fu, Parastoo Sadeghi and Muriel M{\'e}dard }
\thanks{A. Fu and P. Sadeghi are with the Research School of Engineering, College of
Engineering and Computer Science, The Australian National
University, Canberra, Australia. Emails:
\{amy.fu;parastoo.sadeghi\}@anu.edu.au.

M. M{\'e}dard is with the Department of Electrical Engineering and
Computer Science, Research Laboratory of Electronics, Massachusetts
Institute of Technology, Cambridge, MA 02139 USA. Email:
medard@mit.edu. }}
\maketitle
\begin{abstract}

In a single hop broadcast packet erasure network, we demonstrate that it is possible to provide multirate packet delivery outside of what is given by the network min-cut. This is achieved by using a deterministic non-block-based network coding scheme, which allows us to sidestep some of the limitations put in place by the block coding model used to determine the network capacity.

Under the network coding scheme we outline, the sender is able to transmit network coded packets above the channel rate of some receivers, while ensuring that they still experience nonzero delivery rates. Interestingly, in this generalised form of asynchronous network coded broadcast, receivers are not required to obtain knowledge of all packets transmitted so far. Instead, causal feedback from the receivers about packet erasures is used by the sender to determine a network coded transmission that will allow at least one, but often multiple receivers, to deliver their next needed packet.

Although the analysis of deterministic coding schemes is generally a difficult problem, by making some approximations we are able to obtain tractable estimates of the receivers' delivery rates, which are shown to match reasonably well with simulation. Using these estimates, we design a fairness algorithm that allocates the sender's resources so all receivers will experience fair delivery rate performance.

\end{abstract}

\input{kdiff_intro.tex}

\input{kdiff_model.tex}
\input{kdiff_amodel.tex}
\input{kdiff_outline.tex}
\input{kdiff_leadertx.tex}
\input{kdiff_mkdiff.tex}
\input{kdiff_mrates.tex}
\input{kdiff_fairness.tex}
\input{kdiff_conclusion.tex}

\bibliographystyle{IEEEtran}
\bibliography{IEEEabrv,bibtex}

\end{document}

%% file: kdiff_intro.tex
\section{Introduction}

It is well established that the information capacity of a multicast network is given by its min-cut \cite{ahlswede1}, a result that has been extended to broadcast packet erasure channels in \cite{dana06,lun06}. This is a definitive result that cannot be violated under the system model that is considered. An important feature of this model is the requirement of block coding: that is, it is assumed that a block code is used and the original information cannot be recovered until the entire block has been received. Although block coding is practical in most situations, it is not the only possible method of encoding. As a result, it is possible that the limitations imposed by block coding do not necessarily apply in general.

\pmed{A related issue is the simultaneity of decoding by the receivers in the network. Block-based codes, including rateless codes (for example, those developed for degraded broadcast channels and erasure channels \cite{shulmanthesis}), impose the condition that all receivers decode all information before moving onto the next block. \footnote{The only difference is that in rateless codes the goal is generally reducing completion time rather than maintaing achievable rates.} While block-based decoding is predominantly assumed or studied, there have been some results on intermediate decoding of packets \cite{sanghavi07,beimel07,martinianthesis}, albeit without considering the logical packet ordering for delivery to the application.

Linear network coding, shown in \cite{li1} to be capable of achieving network capacity, also provides some interesting counterexamples \cite{sundararajan2,sundararajan3} to the necessity of block coding restrictions. Although a large number of \emph{block-based} network coding schemes exist \cite{swapna10,eryilmaz08,lucani2,gkantsidis2,ho1,park1,chou2,sadeghi1,sorour1,sadeghi3,li11,sorour10,zeng12} which enforce the complete reception of each block, there are also some non-block-based \emph{deterministic} network coding schemes \cite{durvy07,sundararajan2,sundararajan3,sundararajan11,barros1,fu3,wu12}. By relying on feedback from the receivers to determine the coded transmissions, and similar to \cite{sanghavi07,beimel07,martinianthesis}, these schemes can in some cases \cite{sundararajan2,sundararajan3,fu3} allow receivers to decode packets of information \emph{prior} to receiving all information transmitted by the sender.}

We consider the potential implications of using deterministic network coding in the case of a broadcast packet erasure network, comprising of a single sender connected to multiple receivers via a heterogeneous single-hop broadcast packet erasure channel. From \cite{dana06,lun08}, the capacity of this network under the block coding requirement is given by the channel rate of the \med{most challenged receiver,} with the highest packet erasure probability, which is the min-cut of the network. As a consequence, even a single poorly-performing receiver can severely degrade \med{the rate at which packets can be delivered in order to the application across the} entire network. So a fundamental question that has not been considered before is:

\vspace{1mm}
\emph{What will happen \med{to the delivery rates and simultaneity of decoding} if the sender's transmission rate \emph{exceeds} the channel rate of one or more receivers?}
\vspace{1mm}

Under the traditional block coding approach, the delivery rate of these receivers would be zero. Using deterministic network coding, however, it may be possible for the receivers to continue decoding new information, despite never being able to fully receive all information that has so far been transmitted by the sender. As a result, it may be possible to maintain nonzero (in-order) packet delivery rates to all receivers -- even if the sender's transmission rate exceeds their channel capacity. We demonstrate the principles behind this somewhat counterintuitive statement using the following example, which is based on the two receiver transmission scheme first presented in \cite{durvy07}.

\subsection{Deterministic coding example}
Imagine a simple broadcast network consisting of a single sender aiming to deliver a set of packets in the correct order to two receivers: a primary receiver $\rx_1$ with channel rate $\mu_1<1$ and a secondary receiver $\rx_2$ with channel rate $0<\mu_2<\mu_1$. At each time slot $t$ the sender transmits a single packet, to which the receivers respond with an acknowledgement if they were able to receive it correctly. The sender uses this feedback to determine which coded combination should be transmitted next, as outlined in Algorithm \ref{alg:tworx}. A simple example of this algorithm in progress is given in Table \ref{tab:tworx}. The sender's priority is to transmit packets required by $\rx_1$, who has delivered the most packets so far, but it will also aid $\rx_2$ wherever possible.

Packet transmissions intended for $\rx_1$ are not immediately useful to $\rx_2$. However $\rx_2$ has the ability to store and make use of any of these packets it overhears. For example, in Table \ref{tab:tworx} at $t=1$, $\rx_2$ stores $\pkt_9$, which was intended for $\rx_1$. If, owing to an erasure at $\rx_1$, the sender retransmits $\pkt_9$, it can take the opportunity to code a second packet, $\pkt_4$, intended for $\rx_2$, into $s(t)$. This results in the time $t=2$ transmission of $s(t)=\pkt_9+\pkt_4$.

The inclusion of $\pkt_4$ into $s(t)$ does not interfere with $\rx_1$'s ability to receive $\pkt_9$, since $\rx_1$ has already decoded all packets that could potentially be required by $\rx_2$. Conversely, $\rx_2$ will be able to decode $\pkt_4$ from the transmission, since it already received $\pkt_9$ in the previous time slot. When receivers' contributions to the transmission $s(t)$ do not impact each others' ability to decode, something we call a \emph{knowledge differential transmission} occurs. These opportunities will be discussed in a more general setting later on. The end result in this example is that the sender will, as a first priority, deliver packets to $\rx_1$, resulting in a delivery rate of $\mu_1$ to that receiver. However there are also occasional opportunities to deliver packets to $\rx_2$, resulting in a small nonzero delivery rate at $\rx_2$. We shall show that this idea can be generalised to $R$ receivers, using the deterministic coding scheme studied in \cite{sundararajan2,fu3}.

In summary, the key idea in this paper is to break away from the block-based and rate limited schemes that are prevalent in the literature, and apply a \emph{feedback-based, deterministic} network coding scheme as shown above. In doing so, we can achieve rates that exceed the limitations on capacity determined under the complete information acquisition requirement \cite{dana06}. In a sense, this encoding scheme relaxes the notion of broadcasting common information. Although it still ensures that all receivers can decode all packets, receivers are permitted to do so at greatly differing times. By taking advantage of the leading receiver's progress and the stored packets in the buffers of lagging receivers, we can establish a more general form of broadcast, which we can call \emph{asynchronous broadcast}, which allows receivers to deliver the same information at different rates.

\begin{algorithm}
\caption{Two-receiver transmission algorithm of \cite{durvy07}.}\label{alg:tworx}
\begin{algorithmic}[1]
\STATE Let $\nnt{i}$ be the first undecoded packet of $\rx_i$.
\IF {$\rx_2$ has already received $\nnt{1}$}
  \STATE transmit the XOR combination $\nnt{1}\oplus\nnt{2}$.
\ELSE
  \STATE transmit $\nnt{1}$.
\ENDIF
\end{algorithmic}
\tablespace
\end{algorithm}

\begin{table} \begin{center}
\begin{tabular}{|c|c|c c|c c|}
\hline $t$ & $s(t)$ & \multicolumn{2}{|c|}{$\rx_1$} &
\multicolumn{2}{|c|}{$\rx_2$} \\
 \hline 0 & & & $\pkt_1$-$\pkt_8$ &&$\pkt_1$-$\pkt_3,\pkt_5$ \\
 \hline 1 & $\pkt_9$ & \cross &$\pkt_1$-$\pkt_8$& \tick &$\pkt_1$-$\pkt_3,\pkt_5,\pkt_9$ \\
 \hline 2 & $\pkt_9+\pkt_4$&\tick &$\pkt_1$-$\pkt_9$ & \tick & $\pkt_1$-$\pkt_5,\pkt_9$\\ \hline
\end{tabular}
\tablespace
\caption{An example of the transmission scheme for two
receivers, showing which packets have been decoded by each receiver.
The notation $\pkt_1$-$\pkt_8$ indicates packets
$\pkt_1,...,\pkt_8$ have been decoded. A $\tick$ means
the transmitted packet $s(t)$ was received; a $\cross$
indicates a packet erasure.} \label{tab:tworx}
\end{center} \end{table}

\subsection{Contributions and related work}

In this paper we will demonstrate that it is possible for a wireless broadcast transmission scheme to achieve multi-rate packet delivery of an ordered set of information to a heterogeneous set of receivers. By using a deterministic, non-block based transmission scheme, we are able to improve upon the capacity region established in \cite{dana06}.

Multi-rate broadcast to heterogeneous receivers has been previously considered in the context of applications such as video broadcast and layered streaming \cite{sundaram05,cover1}, where different subsets of the information are made available to different receivers. To our knowledge, the transmission scheme we will present is the first to enable the multi-rate delivery of the same set of ordered information to all receivers.

Our transmission scheme is based on a network coding scheme first proposed in \cite{sundararajan2}, which aimed to minimise the in-order packet delivery delay of a homogeneous network. By making some minor modifications to this scheme, we show that it can be adapted for transmission in a heterogeneous broadcast network. Using an analytical model to approximate the system behaviour, we demonstrate that the delivery rate of this transmission scheme can be accurately estimated. The analysis of feedback-based deterministic coding schemes is generally a challenging problem, owing to the number of cross-receiver dependencies that must be considered. We believe that our analysis is the first attempt at this level of detail.


When implementing multi-rate broadcast, one must also consider how resources are allocated amongst receivers. \med{We therefore introduce a measure of \emph{fairness} whereby a transmission scheme is considered fair if the receivers' in-order packet delivery rates are proportional to their channel rate.} To enable the sender to provide fair multi-rate delivery to the receivers, we introduce the concept of time sharing among different \emph{modes}. Using a fairness algorithm to adjust the amount of time spent in each mode, we illustrate how the sender's resources can be allocated in our transmission scheme to allow \med{good} delivery rates to be achieved by all receivers \pmed{under the aforementioned fairness constraint}.

%% file: kdiff_model.tex
\section{Model}\label{sec:model}
A single sender aims to transmit a set of backlogged packets $\pkt_1,\pkt_2,\med{\pkt_3},...$ in the correct order to a set of $R$ receivers, $\rx_1,...,\rx_R$.\footnote{For our analysis, we assume there is an infinite number of packets to be delivered.} \med{Time is slotted, and the sender transmits at a rate of one packet per time slot.} The channel between the sender and receivers is a wireless broadcast \med{packet} erasure channel, \pmed{where erasures are assumed to occur according to an ergodic Bernoulli random process. More specifically, in each time slot, each receiver $\rx_i$ experiences an independent packet erasure with a probability $\mu_i$, which is called its \emph{channel rate}, as illustrated in Fig. \ref{fig:txscheme}, so that on average $\mu_i$ of the broadcasted packets are successfully received. Similar erasure channel models have been considered in \cite{dana06,sundararajan2,sundararajan3}.} We assume that the receivers' channel rates are unique, and order the receivers so that $\mu_1>\mu_2>...>\mu_R$, so that $\rx_1$ is the \med{\emph{fastest}} receiver with the highest channel rate, followed by $\rx_2$, $\rx_3$ and so on. For simplicity of notation, the set of channel rates is represented by the vector $\muv=[\mu_1,\mu_2,...,\mu_R]$. The \emph{receiver index} $i$ of a receiver $\rx_i$ is therefore a numerical ranking of its channel rate, compared to all other receivers in the system.

At each time slot $t$ the sender broadcasts a single network coded packet \med{$s(t)$, along with the vector of coding coefficients used. The network coding coefficients are chosen from the Galois field $\mathbb{F}_M$, where the field size $M \geq R$, typically the smallest power of 2 satisfying this requirement.} \pmed{For simplicity, we treat each packet $\pkt_n$ as a single symbol from $\mathbb{F}_M$.\footnote{In practice, a packet $\pkt_n$ would consist of a vector of $k$ such symbols, with $\pkt_n \in \mathbb{F}_M^k$, as presented in work such as \cite{chou2,sundararajan11,fragouli1}, with the same coding operations applied to each element of the vector.}}

Each receiver stores the packets it receives, along with the corresponding coding coefficients, in its own buffer.\footnote{It should be noted that under our transmission scheme, it is necessary for all receivers to have enough space to store all packets being coded by the sender.} To decode the original packets, receivers \pmed{use the information from the coding coefficients to apply Gaussian elimination \cite{artin91,koetter1} operations, which, once enough information has been received,} allow them recover the original uncoded packets. After each packet transmission, the receiver sends an acknowledgement if the packet was successfully received, or a negative acknowledgement if the packet was discarded due to an erasure. For the purposes of this work we assume the sender detects these acknowledgements without delay or error. The sender uses this information to record which packets each receiver has stored in its buffer. Based on this information, a \emph{transmission scheme} is used to determine the packet combinations the sender will transmit. The components of the transmission schemes we will study are now outlined.

\subsection{Transmission scheme}
\begin{figure*}
\begin{center}
\begin{tikzpicture}[auto]
 \tikzstyle{block} = [rectangle, draw, text width=6em, text centered,minimum height=3em,node distance=3cm]
 \tikzstyle{block2} = [rectangle, draw, text width=5em, text centered,minimum height=1.8em,node distance=0.75cm]
 \tikzstyle{box}=[rectangle,draw,minimum height=4em,text width=15em, text top]
 \tikzstyle{line} = [draw, -latex']
 \node[block](backlog)[]{Packet backlog};
 \node[block](txsubset)[right of=backlog]{Mode selection};
 \node[block](cod)[right of=txsubset]{Coding scheme};
 \node[block2](rx2)[right of=cod,node distance =4.5cm]{$\cdots$};
 \node[block2](rx3)[below of=rx2]{Receiver $R$};
 \node[block2](rx1)[above of=rx2]{Receiver 1};
 \path [line] (backlog) --  (txsubset);
 \path [line] (txsubset) -- (cod);
 \path [line, dashed](cod) edge node[rectangle] {} (rx2);
 \path [line,dashed] (cod)-- node[above,align=center, text width=8em] {Wireless packet erasure channels\\ $\;$\\\red{$\mu_1$}} (rx1);
 \path [line,dashed] (cod)--node[below] {\red{$\mu_R$}}(rx3);

\end{tikzpicture}
 \caption{A block diagram of the components of \red{the} transmission scheme.} \label{fig:txscheme}
\end{center}
\end{figure*}
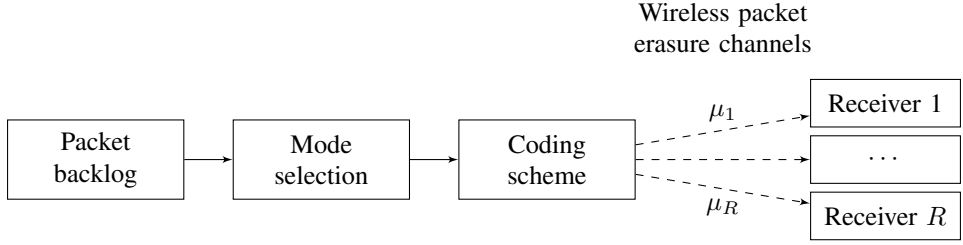

Here we give an overview of the \blue{transmission scheme} we will be studying
in this paper. As shown in Fig. \ref{fig:txscheme} the transmission scheme consists of three components:
\begin{itemize}
 \item A \emph{packet backlog} which contains all packets not yet \blue{\emph{delivered} (to be defined shortly)} to all receivers,
 \item A \emph{mode selector} which determines in which \emph{mode} $m$ (to be defined later) the coding scheme transmits,
 \item A \emph{coding scheme} which determines what network coded combination of the packets in the sender's packet backlog will be selected for transmission.
\end{itemize}

The aim of the transmission scheme is to deliver the packets in order, though not necessarily at the same rate, to all of the receivers.

\pmed{\begin{definition}
A \emph{packet}, $\pkt_n$, corresponds to the $n\ith$ packet in the sender's backlog, and is represented by a symbol in $\mathbb{F}_M$.
\end{definition}}

\pmed{\begin{definition}
A \emph{transmission $s(t)$} is a symbol in $\mathbb{F}_M$ that represents the network coded packet transmitted by the sender at time $t$,
\beq s(t)=\sum_i \alpha_i(t) \pkt_i, \eeq
where the coding coefficients $\alpha_1(t),\alpha_2,\cdots$ are symbols chosen from the field $\mathbb{F}_M$.
\end{definition}}


\med{\begin{definition}
A packet $\pkt_n$ is \emph{decoded} by a receiver $\rx_i$ if Gaussian elimination on all transmissions $s(t)$ it has received so far reveals the value of $\pkt_n$.
\end{definition}}

\begin{definition}
A packet $\pkt_n$ is \emph{older} than another packet $\pkt_m$ if its packet index $n<m$, and \emph{newer} if its packet index $n$ is higher.
\end{definition}
\med{The packets in the sender's backlog are ordered from oldest to newest, so packets are delivered in order from oldest to newest.}

\begin{definition} \label{def:deliv}
A packet $\pkt_n$ is \emph{delivered} to a receiver $\rx_i$ if all \med{older} packets $\pkt_1,...,\pkt_n$ have been decoded by that receiver. $\deliv_i(t)$ is the number of packets delivered to $\rx_i$ by time $t$, so the most recently delivered packet is $\pkt_{\deliv_i(t)}$.
\end{definition}

\begin{definition}
The \emph{delivery rate} for $\rx_i$ is given by $R_i=\deliv_i(t)/t$, the average rate at which packets are delivered to the receiver, as $t \rightarrow \infty$.
\begin{example}
In Table \ref{tab:tworx} at $t=0$, packets $\pkt_1,...,\pkt_8$ have been delivered to $\rx_1$, while $\pkt_1,...,\pkt_3$ have been delivered to $\rx_2$.
\end{example}
\end{definition}

\begin{definition} \label{def:nextneeded}
At any time $t$, the \emph{next needed} packet $\nnt{i}$ of a receiver $\rx_i$ is the oldest packet which has not been delivered to that receiver's buffer. \med{From Definition \ref{def:deliv}, $\nnt{i}=\pkt_{\deliv_i(t)+1}$.}
\begin{example}
In Table \ref{tab:tx}, at time $t=0$ the next needed packets of each receiver are $\nn{1}{0}=\pkt_{11}$, $\nn{2}{0}=\pkt_{5}$ and $\nn{3}{0}=\pkt_{3}$.  
\end{example}
\end{definition}

\subsection{Mode selection}

\begin{definition}
In \emph{mode $m$}, a transmission is restricted to coding packets required by receivers $\rx_m,...,\rx_R$. These receivers are said to be the \emph{mode $m$ receivers}. \red{A transmission made while the sender is in mode $m$ is called a \emph{mode $m$ transmission}.}
\end{definition}
\begin{definition}\label{def:beta}
The \emph{mode vector} $\betav=[\beta_1,...,\beta_R]$ determines the proportion of time spent in each mode. $\beta_m$ is the proportion of time spent in mode $m$, and $\sum_{m=1}^R\beta_m=1$.
\end{definition}
At each time slot, the sender randomly chooses a mode to transmit in, selecting transmission mode $m$ with probability $\beta_m$ \med{independently of previous transmission mode choices.}

%
%

\subsection{Coding scheme}\label{sec:codscheme}
The coding scheme we will study is an adaptation of the coding scheme proposed in \cite{sundararajan2}.\footnote{\med{There are two main differences with \cite{sundararajan2}. Firstly, in this paper, rather than \blue{determining} packet encodings according to \blue{a Bernoulli} arrival pattern as illustrated in \cite{sundararajan2}, we use modes to determine which receivers' packets should be coded. Secondly, in \cite{sundararajan2} it is suggested that any field element may be chosen, but in our implementation, we choose the smallest coefficients that will allow innovative transmissions.}} The method for determining each transmission $s(t)$ is given in Algorithm \ref{alg:codscheme}.\footnote{It can be shown that the \emph{veto list} concept used in \cite{sundararajan2} only impacts the selection of the current coefficient, therefore Algorithm \ref{alg:codscheme} presents a slightly simpler but equivalent form of the coding scheme in \cite{sundararajan2}.} Under this coding scheme, the sender is restricted to coding from the next needed packets of the mode $m$ receivers, so that
 \beq s(t)=\sum_{i=m}^R \alpha_i(t) \nnt{i}. \eeq
Under Algorithm \ref{alg:codscheme}, the sender makes a list containing each oldest undelivered packet $\pkt_j$ and the corresponding set of receivers $G_j$. Then, beginning with the newest packet in the list, it adds each packet $\pkt_j$ into $s(t)$ only if the receiver(s) in $G_j$ would not otherwise receive an innovative packet. 

In Algorithm \ref{alg:codscheme}, lines 13-15 ensure that a packet $\pkt_j$ is only coded into $s(t)$ if $0 \in v_j$. From lines 7-8, this is contingent on at least one of the receivers in $G_j$ being able to cancel out the existing packets in $s(t)$ using Gaussian elimination. Therefore Algorithm \ref{alg:codscheme} \blue{has the} following property.

\begin{property}\label{pr:txcod}
In mode $m$, only the next needed packets $\nnt{m},...,\nnt{R}$ of receivers $\rx_m,...,\rx_R$ can be coded into the transmission $s(t)$.
Of these, a packet $\nnt{i}$ will only be coded into $s(t)$ if it allows $\rx_i$ to deliver its next needed packet.
\end{property}

\begin{example}

\blue{An example of Algorithm \ref{alg:codscheme} is given in Table \ref{tab:tx} for the case where each group $G_j$ consists of a single receiver. Consider $t = 2$. The coding scheme starts \red{by setting} $s(2) = \pkt_{11}$. However, $\rx_2$ can already cancel out this transmission using $\pkt_{11}$ in its buffer ($0 \in v_2$ in line 13 of Algorithm \ref{alg:codscheme}). Hence, $s(2) = \pkt_{11} + \pkt_5$, because $N_2(2) = \pkt_5$. This is already innovative for $\rx_3$. Therefore, $\pkt_3$ need not be added.}
\end{example}

In \cite{sundararajan2} it is shown that, using this coding scheme, we can always find a transmission that
is innovative for all receivers, if the field size $M \geq R$. It should be noted however that while every packet is innovative, receivers will still experience different delivery rates, since the packets they receive are not always contiguous, \med{and, from Definition \ref{def:deliv}, delivery can only take place if all older packets have been decoded at that receiver.}

\begin{algorithm}
\begin{algorithmic}[1]
\STATE Organise receivers $\rx_m,\cdots,\rx_R$ into groups $G_j$, so that $G_j$ contains all receivers
$\rx_i$ whose next needed packet $N_i(t)=\pkt_j$. Let $G$ be the set of all groups $G_j$ which
contain at least one receiver. \STATE Initialise $s(t)=0$. \FOR {each group $G_j \in G$, from high
to low $j$,}
  \STATE Initialise the empty \emph{veto list} $v_j=\left\{\right\}$.
  \FOR {each receiver $\rx_i \in G_j$}
    \STATE Calculate $r_i$, the result of performing Gaussian elimination on $s(t)$ with the \med{transmissions stored} in $\rx_i$'s buffer.
    \IF {$r_i=0$}
       \STATE $v_j \leftarrow v_j \cup \{0\}$.
    \ELSIF {$r_i=\alpha \pkt_j$ for some field element $\alpha$}
      \STATE $v_j \leftarrow v_j \cup \{\alpha\}$.
    \ENDIF
  \ENDFOR
  \IF {$0 \in v_j$}
   \STATE \med{$a_j \triangleq \min(\mathbb{F}_M \backslash v_j)$.} 
   \STATE Set $s(t)=s(t)+a_j\pkt_j$.
  \ENDIF
\ENDFOR
\end{algorithmic}
\tablespace
\caption{Coding scheme.} \label{alg:codscheme}
\end{algorithm}

%% file: kdiff_amodel.tex
\section{Analytical model} \label{sec:amodel}
It is difficult to model the coding scheme presented in Section \ref{sec:model} precisely. There is a complicated set of interdependencies between the packets transmitted by the sender and information stored in the receivers' buffers. The sender's transmissions are determined by the contents of the receivers' buffers, which is in turn determined by a combination of past transmissions, the receivers' erasure patterns and the effects of Gaussian elimination. These interdependencies makes this system difficult to model precisely. Therefore, in order to obtain a tractable model, we make some simplifying approximations in our delivery rate analysis. Here we list further approximations we will use in this paper to obtain a tractable model for calculating the receivers' delivery rates under our transmission scheme. In Sections \ref{sec:Lcalc} to \ref{sec:mrates} we will demonstrate that the numerical results obtained under this analytical model match well with simulations.

\subsection{Simplified encoding scheme} \label{sec:simplecod}

\med{We now analyse Algorithm \ref{alg:codscheme} and} show that the majority of the time the sender implements a simplified version of the coding scheme, 
\med{given in Algorithm \ref{alg:simplecod}}. Where the conditions required to use this simplified coding scheme are satisfied, we can prove that a field size $M=2$ is sufficient to guarantee that \blue{a sender transmission $s(t)$ which is innovative for all mode $m$ receivers can always be found}.

\begin{approximation} \label{ap:heterogrates}
For the transmission schemes studied in this paper, if the receivers' channel rates are distinct and ordered so that $\mu_1>\mu_2>\cdots>\mu_R$, then their delivery rates will also obey the ordering $R_1>R_2>\cdots>R_R$.
\end{approximation}
Intuitively it makes sense that higher channel rates will result in higher packet delivery rates
for the \red{corresponding} receivers. Where Approximation
\ref{ap:heterogrates} holds, \med{the total number of packets delivered by each of the receivers will diverge over time. Therefore,} after a sufficient amount of time has passed the next needed packets $\nnt{1},\cdots,\nnt{R}$ of each receiver will also be distinct with high probability.\footnote{Note
however that this is not always the case: for example at $t=0$, $\nnt{1}=\cdots=\nnt{R}=\pkt_1$ since
none of the receivers has received any transmissions yet.} Therefore we state the next
approximation that we will use for this paper,

\begin{approximation}\label{ap:orderednnt}
The next needed packet of each receiver is distinct and ordered by receiver
index, so that $\nnt{1}>\nnt{2}>\cdots>\nnt{R}$.
\end{approximation}

Under Approximation \ref{ap:orderednnt}, each receiver's next needed packet is distinct. When this is the case, \blue{there is only a single receiver in group $G_j$ corresponding to each next needed packet $\pkt_j$ and} Algorithm \ref{alg:codscheme} reduces to the simplified encoding scheme presented in Algorithm \ref{alg:simplecod}. This \pmed{simplified encoding scheme, which is based on the much smaller field size $\mathbb{F}_2$,} will form the basis of our analysis.

\begin{algorithm}
\caption{Simplified encoding scheme.} \label{alg:simplecod}
\begin{algorithmic}[1]
\STATE Initialise $s(t)=0$.
\FOR {$i=1,\cdots,R$}
\STATE Calculate $r_i$, the result of performing Gaussian elimination on $s(t)$ using the packets in receiver $\rx_i$'s buffer.
\IF {$r_i=0$}
  \STATE $s(t)=s(t)\oplus\nnt{i}$.
\ENDIF
\ENDFOR
\end{algorithmic}
\tablespace
\end{algorithm}

\begin{property}\label{pr:minnov}
Under Algorithm \ref{alg:simplecod}, a mode $m$ transmission is innovative to every mode $m$ receiver, and non-innovative to
receivers $\rx_1,\cdots,\rx_{m-1}$.
\begin{IEEEproof}\upshape{
Under Approximation \ref{ap:orderednnt}, $\rx_1$ will have the newest next needed packet, followed by $\rx_2$, $\rx_3$, and so on. At each iteration $i$ of the \textbf{for} loop, it is ensured that $s(t)$ is innovative to $\rx_i$. By Definition \ref{def:nextneeded} and Approximation \ref{ap:orderednnt}, $\nnt{i+1},\cdots,\nnt{R}$ have already been decoded by $\rx_i$, so the addition of these packets to $s(t)$ in later iterations of the for loop does not affect the innovativeness of $s(t)$ to $\rx_i$. Algorithm \ref{alg:simplecod} guarantees a transmission $s(t)$ will be found which is innovative to all receivers $\rx_m,\cdots,\rx_R$.

If the ordering of next needed packets is maintained, packets coded for
$\rx_m,\cdots,\rx_R$ are already seen by $\rx_1,\cdots\rx_{m-1}$. Therefore the $m^{\mbox{th}}$ mode
will not contain innovative information for receivers $\rx_1$ to $\rx_{m-1}$.}
\end{IEEEproof}
\end{property}

The simplified transmission scheme of Algorithm \ref{alg:simplecod} also has the following property:
\begin{property} \label{pr:nntGF2}
Implementing Algorithm \ref{alg:simplecod} with the binary field $\mathbb{F}_2$ is sufficient to ensure that Property \ref{pr:minnov} holds.
\begin{IEEEproof}\upshape{
Under Approximation \ref{ap:orderednnt}, the next needed packets of each receiver are distinct. Therefore each group $G_j \in G$ contains exactly one receiver. It follows that in each iteration of the \textbf{for} loop corresponding to a group $G_j \in G$, only one element can be added to the corresponding veto list $v_j$. \blue{Since the binary field $\mathbb{F}_2$ contains two elements, this field is sufficient to ensure an appropriate coefficient can always be found.}}
\end{IEEEproof}
\end{property}

An example of the simplified transmission scheme for three receivers \red{in mode 1} is given in Table \ref{tab:tx}. Note that each packet coded into $s(t)$ corresponds to a receiver being able to decode its next needed packet. \med{Two important features of this coding scheme can be observed:}
\begin{enumerate}
    \item It is possible for packets to be delivered to any of the receivers. For example, at time $t=2$, $\rx_2$ delivers its next needed packet $\pkt_5$, and at $t=4$ receivers $\rx_1, \rx_2, \rx_3$ all deliver their next needed packets $\pkt_{11}, \pkt_7$ and $\pkt_3$, respectively.
    \item Every next needed packet coded into a transmission $s(t)$ corresponds to a receiver who can decode \blue{and hence deliver} its next needed packet, if they \red{receive} the transmission. For example, at time $t=4$, all three receivers' next needed packets are encoded. Therefore, receiving the transmission allows all three receivers to \blue{deliver} in the same time slot.
\end{enumerate}


\begin{table*}[ht] \begin{center}
\begin{tabular}{|c|c|c c|c c|c c|}
\hline $t$ & $s(t)$ & \multicolumn{2}{|c|}{$\rx_1$} &
\multicolumn{2}{|c|}{$\rx_2$} & \multicolumn{2}{|c|}{$\rx_3$}\\

\hline 0& & &$\pkt_1$-$\pkt_{10}$& &$\pkt_1$-$\pkt_4,\pkt_6$& &$\pkt_1$-$\pkt_2,\pkt_7$\\
\hline 1&$\pkt_{11}$&\cross&$\pkt_1$-$\pkt_{10}$&\tick&$\pkt_1$-$\pkt_4,\pkt_6,\pkt_{11}$&\cross&$\pkt_1$-$\pkt_2,\pkt_7$\\
\hline 2&$\pkt_{11}+\pkt_5$&\cross&$\pkt_1$-$\pkt_{10}$&\tick&$\pkt_1$-$\pkt_6,\pkt_{11}$&\tick&$\pkt_1$-$\pkt_2,\pkt_7,\pkt_{11}+\pkt_5$\\
\hline 3&$\pkt_{11}+\pkt_7$&\cross&$\pkt_1$-$\pkt_{10}$&\cross&$\pkt_1$-$\pkt_6,\pkt_{11}$&\tick&$\pkt_1$-$\pkt_2,\pkt_5,\pkt_7,\pkt_{11}$\\
\hline 4&$\pkt_{11}+\pkt_7+\pkt_3$&\tick&$\pkt_1$-$\pkt_{11}$&\tick&$\pkt_1$-$\pkt_7,\pkt_{11}$&\tick&$\pkt_1$-$\pkt_3,\pkt_7,\pkt_{11}$\\
\hline
\end{tabular}
\tablespace
\caption{An example of mode 1 transmission, and the effect on receivers'
buffers. Ticks and crosses represent receivers' channel states \blue{(successful and unsuccessful reception, respectively)} at each
transmission.} \label{tab:tx}

\end{center} \end{table*}

\subsection{Buffer model} \label{sec:mod:buffer}

We outline the model we will use in our analysis of the receivers' buffer's contents. The packets stored in the buffer serve a dual purpose. Firstly, they provide information that will facilitate the decoding of the received packets. Secondly, they store information about packets that will need to be delivered at some point in the future.

For the purpose of analysis, we can imagine that every receiver's buffer has space allocated for all of the packets in the sender's backlog. \med{This can be imagined \pmed{as a large vector, with each element corresponding to a packet,} ordered left to right from oldest to newest.} \pmed{The associated coding coefficients are also stored along with each element of the vector, so that the receiver has the information required to perform Gaussian elimination.} As illustrated in Fig. \ref{fig:buffer}, each position $n$ in the buffer corresponds to the packet $\pkt_n$. \red{The following will be used to describe the contents of a receiver's buffer.}

\begin{definition}
\pmed{For a receiver $\rx_i$, The \emph{buffer coefficient matrix} $\bufpkt_i$ stores the coding coefficients associated with each buffer position. Each packet $\pkt_n$ corresponds to the $n\ith$ column of the buffer coefficient matrix, which we denote by $\bufpkt_i(n)$. If packet $\pkt_i$ is coded into the combination stored at $\pkt_n$, then the $i\ith$ element of that column vector will correspond to its coding coefficient. Otherwise, this element will be 0. If buffer position $n$ is empty, then $\bufpkt_i(n)=\mathbf{0}$.}
\end{definition}

\pmed{To decode the packets stored in the buffer, identical Gaussian elimination operations are applied to the columns of the buffer coefficient matrix, and the packets in the buffer. Once a column of the buffer coefficient matrix $\bufpkt_i(n)$ has been reduced to a standard basis vector $e_n$, then the corresponding packet has been decoded.}

\begin{definition}
\blue{For a receiver $\rx_i$, the} \emph{buffer fill vector} $\buf_i$ indicates whether any information has been stored at each position $n$ \pmed{of the buffer}. We say that $\buf_i(n)=0$ if \pmed{$\bufpkt_i(n)$ is the zero vector, otherwise} $\buf_i(n)=1$.
\end{definition}

Generally, the receiver will be required to store both coded and uncoded packets. Algorithm \ref{alg:buf} outlines procedure for storing a received transmission $s(t)$ in the buffer. This procedure can be divided into three stages: using Gaussian elimination to reduce $s(t)$, storing the result in the correct buffer position, and using this new packet to further reduce other packets in the buffer. We now give an example of Algorithm \ref{alg:buf} at work.
\begin{example}
In Table \ref{tab:tx}, at time $t=2$, $\rx_3$ cannot further reduce $s(2)=\pkt_{11}+\pkt_5$, nor can $s(2)$ further reduce any of the packets stored by $\rx_3$. Therefore the received packet $\pkt_{11}+\pkt_5$ will be stored at buffer position 11. By contrast, at time $t=3$ the presence of packets $\pkt_7$ and $\pkt_{11}+\pkt_5$ allows the received packet $s(3)=\pkt_{11}+\pkt_7$ to be reduced down to $s'(3)=\pkt_5$. \blue{\red{After} storing $s'(3)=\pkt_5$ at buffer position 5, we use it to reduce $\pkt_{11}+\pkt_{5}$ down to $\pkt_{11}$ which will replace the content of buffer position 11.}
\end{example}

\begin{algorithm}
\caption{Adding a received packet to the buffer of $\rx_i$.}\label{alg:buf}
\begin{algorithmic}[1]
\STATE Find $s'(t)$, the remainder when Gaussian elimination is applied to $s(t)$ using \med{the columns of $\bufpkt_i$}.
\IF {$s'(t)\neq 0$}
  \STATE \red{Add $s'(t)$ to buffer position $n$, where $n$ is the highest packet index in $s'(t)$.}
  \STATE \med{Use $s'(t)$ to apply Gaussian elimination \pmed{to the non-empty buffer positions}, to obtain column echelon form \pmed{for $\bufpkt_i$.}}
\ENDIF
\end{algorithmic}
\end{algorithm}

\begin{property}\label{pr:bufpkts}
\red{For every receiver $\rx_i$, Algorithm \ref{alg:buf} ensures that for every \med{\pmed{filled element $n$ in the buffer}, $\pkt_n$ is the packet with the highest index coded.}}
\begin{IEEEproof}\upshape{
\red{Algorithm \ref{alg:buf} only modifies the contents of the buffer at lines 3 and 4. At line 3, $s'(t)$ is stored at the correct position. Since Gaussian elimination has already been applied to $s'(t)$, applying line 4 will not change the highest packet index of any of the stored combinations. }
}\end{IEEEproof}
\end{property}

We now introduce some terminology that will be useful later.

\begin{definition}\label{def:seen}
A packet $\pkt_n$ is \emph{seen} to receiver $\rx_i$ if $\buf_i(n)=1$, i.e. there is a packet stored at buffer position $n$.\footnote{\med{In this paper we use the definition of \emph{seen} introduced in \cite{sundararajan2}. Note that this differs from the definition used in \cite{sundararajan3,sundararajan11}.}}

\begin{example}
In Table \ref{tab:tx}, at time $t=2$, receiver $\rx_3$ has packets $\pkt_1,\pkt_2,\pkt_7,\pkt_{11}+\pkt_5$ in its buffer. In our model of the buffer, positions 1, 2, 7 and 11 are filled, therefore $\buf_3(1)=\buf_3(2)=\buf_3(7)=\buf_3(11)=1$ \red{and packets $\pkt_1$, $\pkt_2$, $\pkt_7$ and $\pkt_{11}$ have been seen}. The \pmed{nonzero columns of the buffer coefficient matrix are given by $\bufpkt_3(1)=e_1$, $\bufpkt_3(2)=e_2$, $\bufpkt_3(7)=e_7$, and $\bufpkt_3(11)=e_{11}+e_5$.}
\end{example}
\end{definition}

\begin{definition}
A transmission $s(t)$ is \emph{innovative} to a receiver $\rx_i$ if $s(t)$ cannot be computed from packets in $\rx_i$'s buffer.
\end{definition}

This method of representing the buffer has the following useful property:

\begin{property}\label{pr:seendeliv}
\red{A receiver $\rx_i$'s next needed packet is always the first unseen packet in its buffer. In other words, its next needed packet is given by $\nnt{i}=\pkt_n$, where $n$ is the first buffer position for which $\buf_i(n)=0$.}
\begin{IEEEproof}\upshape{
If \red{$\buf_i(1)=\cdots=\buf_i(n-1)=1$}, then packets $\pkt_1,\cdots,\pkt_{n-1}$ must have been decoded and delivered to the receiver \red{(since we have received $n-1$ innovative combinations of $n-1$ packets). By Property \ref{pr:bufpkts}, if $\pkt_n$ has been delivered by $\rx_i$, then we must have $B_i(n)=1$. Since $B_i(n)=0$, it follows that the first unseen packet $\pkt_n$ is the next needed packet of $\rx_i$.}}
\end{IEEEproof}
\end{property}

\subsubsection{Regions of the buffer} \label{subsec:bufregions}
\begin{figure}[width=\columnwidth]\begin{center}
\begin{tikzpicture}
\drawrxbox{1}{6}
\drawlabelbox{\boxwidth}{$\pkt_1$}
\drawlabelbox{\boxwidth*2}{$\pkt_2$}
\drawlabelbox{\boxwidth*4}{$\cdots$}
\drawlabelbox{\boxwidth*7}{$\nnt{i}$}
\drawlabelbox{\boxwidth*15}{$\nnt{1}$}
\drawpatternbox{8}{15}{black!50}
\drawmtbox{16}{18}
\node at (19.5*\boxwidth,0){$\cdots$};
\belowbrace{0.5*\boxwidth}{6.5*\boxwidth}{Delivered}
\belowbrace{7.5*\boxwidth}{14.5*\boxwidth}{Packet density}
\belowbrace{15.5*\boxwidth}{20*\boxwidth}{Empty}
\end{tikzpicture}
\caption{Our model of receiver $\rx_i$'s buffer. Dark coloured packets have been \blue{all received and delivered}; some fraction of the shaded packets have been received; white packets have not been received. Packets are ordered left to right by ascending packet index.}\label{fig:buffer}
\end{center}\end{figure}
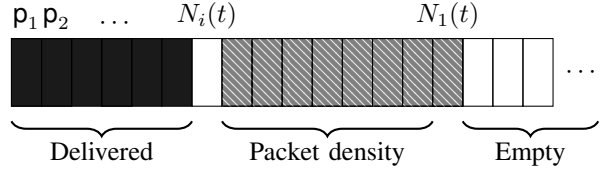

As illustrated in Fig. \ref{fig:buffer}, our model of the buffer of a receiver $\rx_i$ can be broken down into several distinct regions. The \emph{delivered region} consists of all packets \med{$\pkt_1,\cdots,\pkt_{\deliv_i(t)}$ which have been} delivered to the receiver. By definition this is immediately followed by $\nnt{i}=\pkt_{\deliv_i(t)+1}$, the oldest undelivered packet in the buffer. Following this is the \emph{packet density region}, consisting of combinations of transmitted but undelivered packets $\pkt_{\deliv_i(t)+2},\cdots,\nnt{1}$. In this region it is possible for $\rx_i$ to overhear both coded and uncoded transmissions intended for faster receivers. All packets lying beyond $\nnt{1}$ form the \emph{empty region} where no packets are yet stored, since no information about them has been broadcasted yet.

For a receiver $\rx_i$, there are three possible outcomes regarding the storage \blue{of} every received packet in the buffer.
\begin{itemize}
 \item If $s(t)$ is not innovative to $\rx_i$, then the packet will be discarded.
 \item If $s(t)$ is innovative and allows $\nnt{i}$ to be decoded, then $s'(t)$ will be stored at position $\deliv_i(t)+1$. This results in the delivery of one or more new packets, so that $\deliv_i(t+1)>\deliv_i(t)$.
 \item If $s(t)$ is innovative but it does not help decode $\nnt{i}$, then the transmission is stored in the packet density region of the buffer according to Algorithm \ref{alg:buf}. 
\end{itemize}

\subsubsection{Subregions}\label{subsec:subregions}
\begin{figure*}[width=2*\columnwidth]\begin{center}
\begin{tikzpicture}

\drawrxbox{-2}{3}
\drawlabelbox{\boxwidth*4}{$\nnt{4}$}
\drawlabelbox{\boxwidth*10}{$\nnt{3}$}
\drawpatternbox{5}{9}{black!80}
\drawlabelbox{\boxwidth*16}{$\nnt{2}$}
\drawpatternbox{10}{15}{black!40}
\drawlabelbox{\boxwidth*21}{$\nnt{1}$}
\drawpatternbox{16}{21}{black!25}
\drawmtbox{22}{26}
\node at (27.5*\boxwidth,0){$\cdots$};

\belowbrace{-2.5*\boxwidth}{3.5*\boxwidth}{Delivered}
\belowbrace{4.5*\boxwidth}{9.5*\boxwidth}{Subregion 4\\(Buffer density $B_4$)}
\belowbrace{10.5*\boxwidth}{15.5*\boxwidth}{Subregion 3}
\belowbrace{16.5*\boxwidth}{20.5*\boxwidth}{Subregion 2}
\belowbrace{21.5*\boxwidth}{28*\boxwidth}{Subregion 1}
\end{tikzpicture}
\caption{Subregions within the packet density region of receiver $\rx_4$'s buffer. Darker colouring corresponds to a higher packet density\red{; white indicates a packet has not been received}. Packets are ordered left to right by ascending index.}\label{fig:mbuffer}
\end{center}\end{figure*}
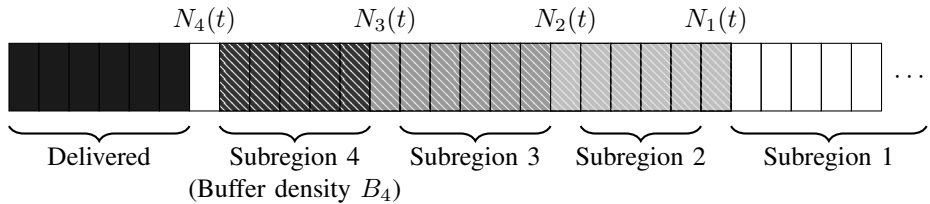

Taking a closer look at the packet density region of Fig. \ref{fig:buffer}, this region can be further divided into subregions $1,2,\cdots,R$, each with its own distinct packet density. This is illustrated in Fig. \ref{fig:mbuffer}. The reason each of these regions exists, is that the transmissions from each mode $m$ contribute to the packet density of each subregion. However, it is not a uniform contribution. Mode 1 transmissions evenly cover packets in subregions $1,2,\cdots,R$, while mode 2 transmissions cover subregions $2,\cdots, R$, and so on. The impact of these transmissions depends on how frequently  they occur, which is in part determined by the mode vector $\betav$. For a given receiver $\rx_r$, the primary subregion of interest is subregion $r$, which gives us the packet density in the subregion where $\rx_r$ is delivering its next needed packets. The packet density of subregion $r$ is called the \emph{buffer density} $B_r$, which will be useful in determining the delivery rate later on.

\begin{definition}\label{def:bufdensity}
For a receiver $\rx_r$, the \emph{buffer density} $B_r$ is defined as the average proportion of buffer positions filled inside subregion $r$,
\beq B_r=\begin{cases}
 0, & r=1\\
 \frac{\sum_{j=\deliv_{r}(t)+2}^{\deliv_{r-1}(t)} \buf_r(j)}{(\deliv_{r-1}(t)-\deliv_{r}(t)-1)}, & 1<r\leq R.
 \end{cases}\eeq
This is illustrated in Fig. \ref{fig:mbuffer}.
\end{definition}

\subsubsection{Buffer model approximations} \label{sec:approx}

In our delivery \red{rate} analysis of the transmission scheme, we model our system for the simple coding scheme presented in Section \ref{sec:simplecod}. However, even \med{to this coding scheme we apply some further simplification for tractability}. Therefore we now present a list of approximations that will be used to model the receivers' buffers.\footnote{It can be observed that all of the approximations in this section are used to remove the more unpredictable elements of cross-receiver interactions. For $\rx_1$ and $\rx_2$, the effective lack of interaction with other receivers means that when analysing these two receivers, our approximations will be completely accurate. As the receiver index increases however, the accuracy of these approximations will slowly decrease.}

\begin{approximation} \label{ap:uncod}
All packets stored in a receiver \red{$\rx_i$'s buffer are uncoded, so if $\buf_i(n)=1$, then $\bufpkt_i(n)=e_n$. This means that once a receiver $\rx_i$ has seen a packet $\pkt_n$, we assume \med{decoding} of it as well}.
\end{approximation}

\med{This approximation results a somewhat optimistic estimate of the probability of decoding. However, this is not an unreasonable approximation as} our coding scheme is \blue{\emph{near-systematic}}, meaning that \red{whenever Property \ref{pr:minnov} permits}, the sender transmits an uncoded packet. And of the coded transmissions received, many will be reduced to uncoded packets when they are stored in the buffer, through Gaussian elimination. From our simulations we observe that, after Gaussian elimination has been applied, the vast majority of packets stored in the buffer are uncoded. Some examples are given in Table \ref{tab:undec}. In general, we observe that a few factors influence the accuracy of Approximation \ref{ap:uncod}. Higher channel rates and a lower receiver index $i$ will reduce the fraction of undecoded packets a receiver $\rx_i$ has stored in its buffer. On the other hand, more time spent in modes $2,\cdots,R$ will reduce the proportion of undecoded packets stored in the receivers' buffers.

\begin{table}\begin{center}
\begin{tabular}{|c|c|c|}
\hline $\muv$ & $\betav$ & Percentage of undecoded packets \\
\hline $[0.8,0.6,0.4,0.2]$ & $[0.85,0.05,0.05,0.05]$ &	$0.0\%, 0.0\%, 2.66\%, 10.26\%$ \\
\hline $[0.9,0.8,\cdots,0.2]$ & $[0.65,0.05,\cdots,0.05]$	& $0.0\%, 0.01\%, 0.63\%, 1.06\%, 2.88 \%, 5.23\%, 11.5\%    16.22\%$\\
\hline $[0.8,0.75,\cdots,0.45]$	& $[1,0,\cdots,0]$ &	$0.0\%, 0.03\%, 0.31\%, 0.60\%,  3.20\%,  6.10\%, 11.93\%, 12.07\%$\\
\hline
\end{tabular}
\tablespace
\caption{Percentage of undecoded packets in each receiver's buffer, under the indicated values of $\muv$ and $\betav$.}\label{tab:undec}
\end{center}\end{table}

\begin{example}
In Table \ref{tab:tx} at time $t=1$, we see a typical situation where the sender sends an uncoded packet, in this case $s(t)=\pkt_{11}$. Although occasionally a coded packet may be stored in the receiver's buffer, for example the buffer of $\rx_3$ at $t=2$, often Gaussian elimination will help decode the packet, illustrated for $\rx_3$ at $t=3$.
\end{example}

\red{Approximation \ref{ap:uncod} leads to the following property, which simplifies our model of the receivers' packet delivery mechanisms. This property will be useful for the knowledge differential calculations of Section \ref{supsec:mkdiff}.}

\begin{property}\label{pr:kdiff}
For a receiver $\rx_r$ to decode its next needed packet $\nnt{r}$, every packet coded into $s(t)$ \red{apart from $\nnt{r}$ must have been seen by $\rx_r$}.
\end{property}

\begin{definition}
The \emph{leader} in a mode $m$ transmission is the receiver $\rx_m$.
\end{definition}
\begin{definition}\label{def:leadtx}
A \emph{leader transmission} of a packet $\pkt_n$ is a transmission $s(t)$ where $\pkt_n$ is the packet \red{with the highest index coded into the transmission, i.e. the next needed packet of the leader. If the sender is in mode $m$ when this transmission occurs, we can more precisely describe it as a \emph{mode $m$ leader transmission}.}
\begin{example}
If the transmission $s(t)=\pkt_3+\pkt_5+\pkt_{10}$, then it is a leader transmission of packet $\pkt_{10}$, the newest packet in the transmission. \red{If the sender was in mode 2 at time $t$, then $s(t)$ is a mode 2 leader transmission of $\pkt_{10}$.}
\end{example}
\end{definition}


\red{We can further simplify our model of the receivers' packet delivery behaviour by limiting the ways in which packets can be seen.}

\begin{approximation}\label{ap:bufpos}
\med{A receiver $\rx_i$ can only have seen an undelivered packet $\pkt_n$ through the reception of a leader transmission of $\pkt_n$.}
\end{approximation}

In practice, due to the effects of Gaussian elimination, it is possible for a receiver $\rx_i$ to see \pmed{an undelivered packet in the packet density region} at any time slot. But for the purpose of analysis, we only allow the receiver to see a packet \pmed{by the most likely method, namely} receiving a leader transmission. As a result of this approximation, we obtain the following property, which will also be useful in Section \ref{supsec:mkdiff}.

\begin{property}\label{pr:kdiffc}
For a packet $\pkt_n$ in a receiver's packet density region to be seen, $\rx_r$ must have received a leader transmission of $\pkt_n$.
\end{property}

\red{One issue this property raises is that, since there is a common sender, the same number of leader transmissions of each packet will be sent to the receivers. As a result, we can expect some correlation between which packets each receiver has seen. These cross receiver dependencies would significantly increase the complexity of our analysis, so for the sake of simplicity we make} the following approximation.

\begin{approximation} \label{ap:indep}
The number of mode $m$ leader transmissions of a packet $\pkt_n$ is independent of the number of mode $b$ leader transmissions of $\pkt_n$, for all $m\neq b$.
\end{approximation}


It can be observed that most of the approximations made in this section are designed to minimise the number of cross-receiver interactions we will need to take into account in our calculations. With the exception of Approximation \ref{ap:orderednnt}, our analytical model is exact for receivers $\rx_1$ and $\rx_2$. But, as the receiver index increases, so does the impact of cross receiver interactions on the receiver's performance. As a result, it is expected that our analytical model will lose some accuracy as the number of receivers increases. Our simulation results indicate that the delivery rates predicted by the analytical model are accurate for $R=4$ and $8$, however the computational complexity of the simulations prevent us from investigating higher values of $R$ with accuracy.

%% file: kdiff_outline.tex
\section{Outline of delivery rate analysis}

\begin{figure}\begin{center}
\begin{tikzpicture}[auto]
\tikzstyle{block} = [rectangle, draw, minimum width=3em, text centered,minimum height=2em,node distance=2.2cm]
\tikzstyle{line} = [draw, -latex']
\node[block](delivrate)[]{$R_r$};
\node[block] (buffer)[above right of=delivrate] {$B_r$};
\node[block](kdiffl)[above left of =buffer]{$K_r^m$};
\node[block](cod)[above left of=kdiffl]{$P^m(C)$};
\node[block](kdiffc)[above right of=cod]{$D_a^b$};

\node[block](L)[above left of=kdiffc]{$L_l(k)$};
\node[block](Lst)[above right of=kdiffc]{$L_l^*(k)$};
\node[block](U)[above of=L]{$U_l$};
\node[block](T)[right of=U]{$T_l(k)$};
\node[block](gam)[right of=Lst]{$\gamma_l$};
\node[block,dashed](beta)[above of=gam] {$\betav$};

\path[line,dashed](beta)--node[above]{$\beta_l$} (T);
\path[line,dashed](beta)-- (gam);
 \node[below left] at ($(beta.south)+(0.05,-0.2)$){$\beta_1...\beta_{l-1}$};

\path [line] (kdiffl) -- node[left]{$K_r^1...K_r^r$} (delivrate);
\path [line] (buffer) -- (delivrate);
\path [line] (cod) -- (kdiffl);
\path [line] (kdiffc) -- (cod);
 \node[above left,align=center] at ($(cod.north east)+(0,0.2)$) {$D_a^b$, for\\ $b<a<r$};
\path [line] (kdiffl)-- (buffer);
 \node[above right] at ($(buffer.north west)+(-0.1,0)$){$K_r^1...K_r^{r-1}$};

\path [line] (kdiffl) [right,bend right=20] edge node{$K_l^1...K_l^{l-1}$} (gam);
\path [line] (gam) edge (T);
\path [line] (gam) edge (Lst);
\path [line] (L)[bend left=20] edge  node[left,align=center]{$L_1(k)$...\\ $L_{l-1}(k)$} (U);
\path [line] (T) -- node{} (L);
\path [line] (U) -- node{} (L);
\path [line] (L) -- node[left,align=center] {$L_1(k)$...$L_{b-1}(k)$} (kdiffc);
\path [line] (Lst) --  node[right,xshift=0.1cm,yshift=0.05cm]{$L_l^*(k)$} (kdiffc);
\path [line] (delivrate)[bend right=20] edge  node[below right]{$R_1,...,R_{r-1}$} (buffer);

\draw[dotted] ($(U.north west)+(-1.4,0.8)$)  rectangle ($(gam.south east)+(0.2,-0.65)$);
 \node[below right] at ($(U.north west)+(-0.7,0.8)$){Section \ref{sec:Lcalc}};
\draw[dotted] ($(U.north west)+(-1.4,-3.65)$)  rectangle ($(gam.south east)+(0.2,-4.8)$);
 \node[above right] at ($(U.north west)+(-0.7,-7.7)$){Section \ref{supsec:mkdiff}};
\draw[dotted] ($(U.north west)+(-1.4,-7.8)$)  rectangle ($(gam.south east)+(0.2,-8)$);
 \node[above right] at ($(U.north west)+(-0.7,-10.9)$){Section \ref{sec:mrates}};

\end{tikzpicture}
\caption{Dependency graph for quantities calculated in Sections \ref{sec:Lcalc} to \ref{sec:mrates}. Labeled arrows indicate which values are required to calculate the next block, while unlabeled arrows indicate the following block depends directly on the value from the previous block. Curved arrows indicate dependencies on previously calculated values.}\label{fig:blockanalysis}
\end{center}\end{figure}
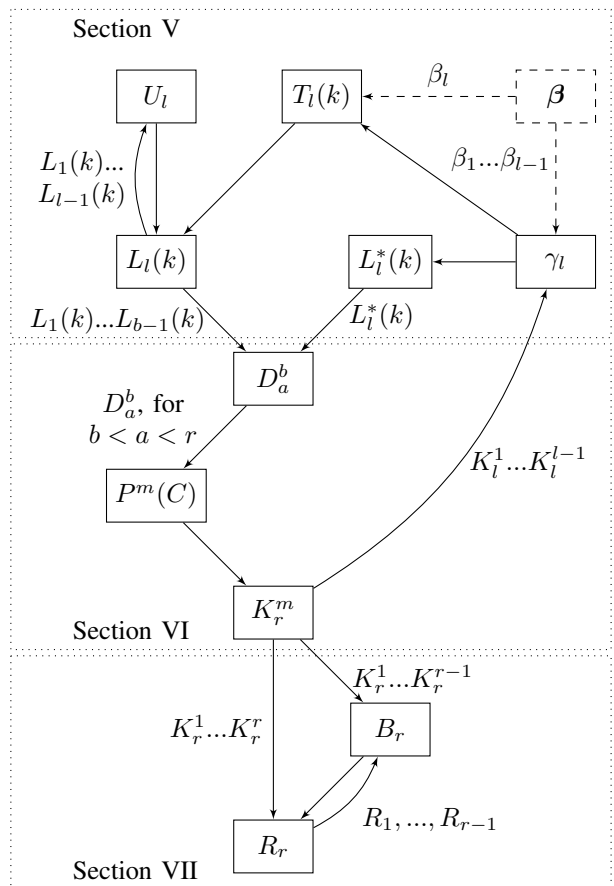

In this section we outline the steps that will be taken in Sections \ref{sec:Lcalc} to \ref{sec:mrates} to determine the delivery rate $R_r$ for each receiver $\rx_r$ under the transmission scheme outlined in Section \ref{sec:codscheme}. The estimation of the delivery rate is a challenging task, given the somewhat complicated dependency between sender transmissions and the contents of each of the receivers' buffers. The approximations made in Section \ref{sec:amodel} go some way toward simplifying these relationships, but the process of calculating the receivers' delivery rates \red{still requires some detail}. A map of the quantities we will calculate on our journey towards finding the delivery rate is illustrated in Fig. \ref{fig:blockanalysis}, with definitions given in Table \ref{tab:defs} for quick reference. The values in this figure will be discussed later, but for now we simply observe that a number of dependencies exist between different components in the calculation. 

\med{To divide the problem into manageable pieces, we make the observation that there are two ways in which a packet may be delivered to a receiver. Either:
\begin{itemize}
 \item The packet was previously seen from leader transmissions of the packet in other modes,
 \item The packet is delivered through a \emph{knowledge differential}, while it is the receiver's next needed packet.
\end{itemize}
The rate analysis is therefore divided into three parts. Analysis of leader transmissions takes place in Section \ref{sec:Lcalc}, while knowledge differentials are studied in Section \ref{supsec:mkdiff}. The results from these two sections are combined to determine the delivery rate in Section \ref{sec:mrates}.
Each of these sections, corresponding to the dotted rectangles in Fig. \ref{fig:blockanalysis}, will now be briefly discussed.}

\begin{table}\begin{center}
\begin{tabular}{|l|p{7cm}|}

\hline $U_l$ & Probability that a packet \med{will not be seen from} $\rx_l$ from mode $1,...,l-1$ transmissions.\\
\hline $T_l(k)$ & Probability that \red{there will be $k$ total mode $l$ leader transmissions of a packet}, given that it was not seen from mode $1,...,l-1$ \red{leader} transmissions.\\
\hline $\betav$ & The mode vector set by the sender (see Definition \ref{def:beta}).\\
\hline $L_l(k)$ & Probability that a total of $k$ mode $l$ leader transmissions were made of a packet that has been delivered by $\rx_l$.\\
\hline $L_l^*(k)$ & Probability that so far, there have been exactly $k$ mode $l$ leader transmissions of $\nnt{l}$, $\rx_l$'s current next needed packet.\\
\hline $\gamma_l$& Probability of a knowledge differential transmission for $\rx_l$. \\
\hline $D_a^b$ & Probability that the next needed packet $\nnt{b}$ of a receiver $\rx_b$ is seen by another receiver $\rx_a$, where $a>b$.\\
\hline $P^m(C)$ & Probability of a coding $C$ being transmitted in mode $m$.\\
\hline $K_r^m$ & Probability that $\rx_r$ has a knowledge differential \red{while the sender is in mode $m$}.\\
\hline $B_r$ & Expected buffer density.\\
\hline $R_r$ & Estimated delivery rate for receiver $\rx_r$.\\
\hline
\end{tabular}
\tablespace
\caption{Definition of the values used in Fig. \ref{fig:blockanalysis}.}\label{tab:defs}
\end{center}\end{table}

\subsection{Leader transmissions}

In Section \ref{sec:Lcalc}, we characterise the number of mode $l$ leader transmissions of a packet $\pkt_n$ in the packet backlog. The process for calculating this probability is illustrated by the top box of Fig. \ref{fig:blockanalysis}. By Approximation \ref{ap:bufpos}, this characterisation is important because it can be used to determine the likelihood that \red{a packet $\pkt_n$ has been seen} by a receiver.

There are three possible cases to consider when finding the likelihood that a packet $\pkt_n$ has been transmitted $k$ times in mode \blue{$l$}. These depend on $\rx_l$'s progress towards delivering $\pkt_n$. \begin{itemize}
 \item If $n\leq \deliv_l(t)$ then $\pkt_n$ has already been delivered by $\rx_l$, and $L_l(k)$, calculated in Section \ref{subsec:Lcalc}, gives the probability \red{there were $k$} \blue{mode $l$ leader transmission\red{s of $\pkt_n$}.}
 \item If $n=\deliv_l(t)+1$, then $\pkt_n=\nnt{l}$, and $k$ takes the probability distribution $L_l^*(k)$, calculated in Section \ref{subsec:Lstcalc}.
 \item If $n>\deliv_l(t)+1$, then there have been no \blue{mode $l$ leader} transmissions of $\pkt_n$ so far, i.e. $k=0$, so we do not need to consider the possibility of any mode $l$ leader transmissions of these packets.
\end{itemize}
The probabilities $L_l(k)$ and $L_l^*(k)$ are calculated in Section \ref{sec:Lcalc}.

\subsection{Knowledge differentials}

In Section \ref{supsec:mkdiff} we find the probability of a knowledge differential. This process is illustrated by the middle box in Fig. \ref{fig:blockanalysis}. Knowledge differentials are a key concept in this paper\med{, as they are necessary to allow receivers to deliver the remaining packets that were not seen from leader transmissions}.
\begin{definition}\label{def:kdiff}
At time $t$, a receiver $\rx_r$ has a \emph{knowledge differential} if the portion of $s(t)$ coded for \blue{faster} receivers $\rx_1,..,\rx_{r-1}$ can be decoded by $\rx_r$ \pmed{using the information stored in its buffer}. \red{If the sender is in mode $m$, then} the probability of a knowledge differential is given by $K_r^m$.

If $s(t)$ meets this criterion, it is a \emph{knowledge differential transmission}. Otherwise, it is a \emph{non-knowledge differential transmission}.
\end{definition}

Knowledge differentials occur when portions of the transmission $s(t)$ intended for \blue{faster} receivers do not affect $\rx_r$'s ability to decode the transmission. Knowledge differentials make it possible to provide different delivery rates to each of the receivers, since some packets will be received effectively without interference from the \blue{faster} receivers. Whenever a knowledge differential \red{transmission is received}, $\rx_r$ is able to decode \blue{and hence deliver its} next needed packet $\nnt{r}$.

\begin{property}\label{ap:nnseen}
\med{A receiver $\rx_i$'s next needed packet $\nnt{i}$ can only be seen through the reception of a knowledge differential transmission.}
\end{property}

Combining Property \ref{pr:txcod} and Approximation \ref{ap:orderednnt}, a new packet can be delivered to $\rx_r$ iff its next needed packet $\nnt{r}$ is encoded into the transmission $s(t)$. In these time slots, the sender can take advantage of the knowledge differential to effectively send $\rx_r$ an uncoded packet.

There are two components required to calculate $K_r^m$: $D_a^b$, the probability that a receiver $\rx_a$ has a $\rx_b$ knowledge difference, and $P^m(C)$, the probability of a \blue{particular} coding $C$. These are now discussed.

\subsubsection{Knowledge differences}
From Property \ref{pr:kdiff}, each of the packets coded into $s(t)$ \blue{for other receivers} must be seen by a receiver $\rx_a$ in order to obtain a knowledge differential. The probability of each of these packets being seen is given by the probability of a $\rx_b$ knowledge difference, defined as follows.

\begin{definition} \label{def:kdiffc}
At time $t$, a receiver $\rx_a$ obtains a \emph{$\rx_b$ knowledge difference} if $\buf_a(\deliv_b(t)+1)=1$, i.e. $\rx_a$ has seen $\nnt{b}$. The probability of $\rx_a$ having a $\rx_b$ knowledge difference is $D_a^b$. If $a\leq b$, then we set $D_a^b=1$.\footnote{We set $D_a^b=1$ when $a \leq b$, since if $a<b$ then $\rx_b$ has already decoded $\nnt{a}$, and when $a=b$, a receiver cannot affect its own probability of decoding.}
\end{definition}

This means that when $\rx_a$ has a $\rx_b$ knowledge difference, \blue{even} if $\nnt{b}$ is coded in the sender transmission $s(t)$, it will not affect $\rx_a$'s ability to decode $s(t)$. Examples of this can be seen in Table \ref{tab:tx}. At times $t=2,3,4$, $\rx_2$ has seen $\rx_1$'s next needed packet, $\pkt_{11}$. At time $t=4$, $\rx_3$ has seen both $\rx_1$ and $\rx_2$'s next needed packets, $\pkt_{11}$ and $\pkt_7$, respectively.

It is important to note that knowledge differences are a common occurrence, since the broadcast nature of the system means that all transmissions may be overheard by any of the receivers. A packet seen by one receiver, but not another, causes a knowledge \red{difference} to occur in the future. Generally the higher a receiver's channel rate, the better its chances of obtaining a knowledge difference, since they are more likely to overhear packets that are missed by other receivers.

\subsubsection{Codings}
The probability of a knowledge differential $K_r^m$ depends on which of the receivers' next needed packets are coded into $s(t)$. Therefore in order to determine the probability of a knowledge differential, it is first necessary to find the probability of every possible coding, $C$.

\begin{definition} \label{def:txcod}
The \emph{transmission coding} $C$ is an $R$-length binary vector representing which of the receivers' next needed packets have been coded into $s(t)$. If the $r\ith$ element of $C$, $C(r)=1$, then $\nnt{r}$ is one of the packets coded into $s(t)$. Otherwise, $C(r)=0$ and $s(t)$ does not code $\nnt{r}$.
\begin{example}
If $R=4$ and $s(t)=\nnt{1}+\nnt{3}+\nnt{4}$, then $C=[1,0,1,1]$. If $R=5$ and \red{$s(t)=\nnt{2}+\nnt{4}$, then $C=[0,1,0,1,0]$}.
\end{example}
\end{definition}

%

\subsection{Delivery rate}

In Section \ref{sec:mrates} we \blue{will} find the delivery rate $R_r$. This process is illustrated in the bottom box in Fig. \ref{fig:blockanalysis}. \pmed{The knowledge differential probabilities established in Section \ref{supsec:mkdiff} are used to determine the delivery rate of all of the receivers as follows.}

\red{By Property \ref{pr:minnov}, every} \blue{mode $l$ transmission} received by a receiver $\rx_r$ \red{is innovative} \blue{for $r \ge l$}. \red{Therefore,} it either allows the delivery of $\nnt{r}$, or is stored in the \red{packet density region}. Knowledge differential transmissions allow delivery of next needed packets, and non-knowledge differential transmissions contribute seen packets to the buffer. The buffer density $B_r$ (\med{recall} Definition \ref{def:bufdensity}) can be determined from the probability of non-knowledge differential transmissions and the delivery rates of receivers $\rx_1,...\rx_{r-1}$.

For a receiver $\rx_r$, the buffer density $B_r$ \pmed{gives} us the fraction of packets which are already seen by $\rx_r$, prior to the contribution of knowledge differential transmissions. The delivery rate can be found by determining how quickly knowledge differential transmissions can fill the remaining fraction of unseen packets.



%% file: kdiff_leadertx.tex
\section{Leader transmissions}\label{sec:Lcalc}

In this section, we find the number of mode $l$ leader transmissions of any packet $\pkt_n$ we can expect, at any time $t$.
All calculations in this section are implicitly dependent on \blue{the mode vector} $\betav$, however for readability we omit direct mention of $\betav$ in our notation. 

\subsection{Mode 1 and 2 leader transmissions}

We begin with a walkthrough of the mode 1 and 2 leader transmissions of an arbitrary packet $\pkt_n$. Recall the definitions of a seen packet (Definition \ref{def:seen}) and a leader transmission (Definition \ref{def:leadtx}).

The first time a packet $\pkt_n$ will be transmitted by the sender is as a mode 1 leader transmission. This will only happen once $\pkt_n=\nnt{1}$, i.e. it is the next needed packet of $\rx_1$. The total number of times $k$ that $\pkt_n$ is transmitted in mode 1 is simply the number of \red{mode 1} transmissions it takes until it \red{is seen} by $\rx_1$.

\begin{example}
Say that at time $t=10$, the receiver $\rx_1$ has delivered $\pkt_1,\cdots,\pkt_7$, so \blue{its} next needed packet is $\nn{1}{10}=\pkt_8$. There are mode 1 transmissions at $t=10,12,13$ and the corresponding $\rx_1$ channel state at these time slots is $\cross,\cross,\tick$. In this case, at times $t=10,11,12$, there will have been $k=1,2,3$ mode 1 leader transmissions of $\pkt_8$ respectively, after which it is seen by $\rx_1$. There are no more mode 1 leader transmissions of $\pkt_8$ once is seen. After this time $\pkt_9$ will become $\rx_1$'s next needed packet, and there will have been a total of $k=3$ mode 1 leader transmissions of $\pkt_8$.
\end{example}

Similarly for mode 2, once $\nnt{2}=\pkt_n$, all mode 2 transmissions will be leader transmissions of $\pkt_n$ until it is seen by $\rx_2$. The difference from the mode 1 case is that some of the earlier mode 1 leader transmissions of a packet $\pkt_n$ may have been overheard by $\rx_2$, therefore there is a probability that $\pkt_n$ is seen \emph{prior} to any mode 2 transmissions. If this is the case, then $k=0$ since \blue{no mode 2 leader \red{transmissions} of $\pkt_n$ \red{are} required.}

\begin{example}
At time $t=11$, the receiver $\rx_2$ has delivered $\pkt_1,\cdots,\pkt_4$ and also has seen $\pkt_6$ and $\pkt_8$. Mode 2 transmissions occur at $t=11,14,17$, and the corresponding $\rx_2$ channel state at these time slots is $\tick,\cross,\tick$. At $t=11$, $\nn{2}{11}=\pkt_5$, and $\rx_2$ receives the first \blue{mode 2} leader transmission of $\pkt_5$. As a result, there is exactly one mode 2 leader transmission of $\pkt_5$. Since $\pkt_6$ is already seen by $\rx_2$, it is immediately delivered and there are no mode 2 leader transmissions of that packet. At time $t=14$, the sender skips straight to $\nn{2}{14}=\pkt_7$. $\rx_2$ experiences an erasure at this time slot, so $\pkt_7$ is transmitted again at $t=17$, giving $k=2$ total mode 2 leader transmissions of $\pkt_7$. Since $\pkt_8$ is already seen, the sender \blue{will skip to $\pkt_9$ as $\rx_2$'s next needed packet \red{the} next time a mode 2 leader transmission occurs.}
\end{example}

From these examples we can make the following observation about mode $l$ leader transmissions.
\begin{property}
There can only be mode $l$ leader transmissions of a packet $\pkt_n$ while it is unseen by $\rx_l$.

\begin{IEEEproof}
\blue{
\upshape{By Property \ref{pr:txcod}, only the receivers' next needed packets can be coded into a transmission. \red{From Property \ref{pr:seendeliv}, the next needed packet of a receiver must always be unseen by that receiver.} It follows that there will be no mode $l$ leader transmissions of a packet $\pkt_n$ that is seen by $\rx_l$.}}
\end{IEEEproof}
\end{property}

\subsection{Buffer position and seen packets}\label{sec:leadtxbuffer}
We now examine the distribution of $k$, the number of mode $l$ leader transmissions, in terms of buffer position. At a given time $t$, a receiver $\rx_l$ has delivered packets $\pkt_1,\cdots,\pkt_{\deliv_l(t)}$. This means that there have also been mode $l$ leader transmissions of some or all of the packets in the corresponding region of the buffer, as illustrated in Table \ref{tab:leadtxs}.

We define the probability $L_l(k)$ of there being $k$ mode $l$ leader transmissions \red{of a given packet $\pkt_n$ for $n\leq\deliv_l(t)$}.  By Approximation \ref{ap:indep} this probability distribution is independent for each value of $l$, so the number of leader transmissions $k$ of a packet $\pkt_n$ in mode $l$ is not correlated to the number of leader transmissions made in other modes.

In contrast, the next needed packet \red{$\nnt{l}=\pkt_{\deliv_l(t)+1}$} has a slightly different probability distribution. Since it is in the process of being delivered, there may or may not have been previous mode $l$ \red{leader} transmissions of this packet. \blue{Therefore, we allow for this packet to have a potentially different probability of $k$ mode $l$ leader transmissions and denote it by $L_l^*(k)$\med{, the probability that there have been exactly $k$ mode $l$ leader transmissions of the packet so far}.}

\begin{table}\begin{center}
\begin{tikzpicture}[auto]
  
  \drawnbrect{0}{-0.4}{1.4}{$\pkt_n=$}{1}
  \drawnbrect{1}{-0.4}{1.4}{$m=1$}{1}
  \drawnbrect{2}{-0.4}{1.4}{$m=2$}{1}
  \drawnbrect{3}{-0.4}{1.4}{$m=3$}{1}
  
  \drawnbrect{0}{1}{1}{$\pkt_1$}{1}
  \drawnbrect{0}{2}{1}{$\cdots$}{1}
  \drawnbrect{0}{3}{1}{$\nnt{3}$}{1}
  \drawnbrect{0}{4}{2}{$\cdots$}{1}
  \drawnbrect{0}{6}{1}{$\nnt{2}$}{1}
  \drawnbrect{0}{7}{1}{$\cdots$}{1}
  \drawnbrect{0}{8}{1}{$\nnt{1}$}{1}
  
  \draw (\rectwidth,-\rectheight) rectangle (\rectwidth*9,-\rectheight*4);
  
  \drawrect{1}{1}{7}{$L_1(k)$}{0.7}
  \drawrect{1}{8}{1}{$L_1^*(k)$}{0.7}
  \drawrect{2}{1}{5}{$L_2(k)$}{0.75}
  \drawrect{2}{6}{1}{$L_2^*(k)$}{0.75}
  \drawrect{3}{1}{2}{$L_3(k)$}{0.84}
  \drawrect{3}{3}{1}{$L_3^*(k)$}{0.84}
\end{tikzpicture}
\tablespace
\caption{A schematic of the three-receiver probability that the sender will transmit $k$ mode $l$ leader transmissions of each packet $\pkt_n$.} \label{tab:leadtxs}
\end{center}\end{table}

\subsubsection{Seen packets}
From this information, we can make some general comments about the methods by which a next needed packet $\nnt{b}$ of a receiver $\rx_b$ can be seen by \blue{another} receiver $\rx_a$, based on its position $n$ in the receiver's buffer. We are primarily interested in studying the next needed packets, since only these packets will be encoded by the sender.

If $b>a$ then $\rx_a$ \red{has already seen} $\nnt{b}$, since by Approximation \ref{ap:orderednnt} it has already been delivered to $\rx_a$. If $a>b$, \red{then by Approximation \ref{ap:bufpos},} the probability of a packet $\nnt{b}$ being seen by another receiver $\rx_a$ depends on the number of leader transmissions of the packet in modes $1,\cdots,b$.\footnote{Modes $b+1,\cdots,R$ do not contribute leader transmissions because $\nnt{b}$ has not yet been transmitted in those modes.} Generally, the more leader transmissions of $\nnt{b}$ there have been, the more likely it is that $\rx_a$ has already seen this packet.


\subsection{Calculating $L_l^*(k)$} \label{subsec:Lstcalc}

We begin with the simpler case: finding the probability $L_l^*(k)$ that $\nnt{l}$, the next needed packet of $\rx_l$, has been transmitted $k$ times so far \red{in mode $l$}.

The sender only transmits mode $l$ leader transmissions of a new packet once the previous next needed packet has been \red{seen} by $\rx_l$. \red{Recall from Approximation \ref{ap:bufpos} that} a next needed packet can be received by two methods:
\begin{itemize}
 \item Mode $l$ \red{leader} transmissions{\footnote{\red{Mode $l$ transmissions are also knowledge differential transmissions, but because we are counting the number of mode $l$ leader transmissions, we treat them separately.}}}
 \item Mode $1,\cdots,l-1$ knowledge differential transmissions.
\end{itemize}
The probability of a mode $l$ transmission is simply $\beta_l$, while the probability of a mode $i=1,\cdots,l-1$ knowledge differential transmission for $\rx_l$ is given by
 \beq \gamma_l=\sum_{m=1}^{l-1} \beta_m K_l^m, \label{eq:gamma}\eeq
where $K_l^i$, the probability that $\rx_l$ has a knowledge differential from a mode $i$ transmission, will be calculated in Section \ref{supsec:mkdiff}.

So from $\rx_l$'s perspective, there are four possible actions from the sender at each time slot\red{, which we now list}. \blue{The sender} can transmit:
\begin{enumerate}
 \item A mode $l$ transmission, with probability $\beta_l$
 \item A mode $1,\cdots,l-1$ knowledge differential transmission, with probability $\gamma_l$
 \item A mode $1,\cdots,l-1$ non-knowledge differential transmission
 \item A mode $l+1,\cdots,R$ transmission
\end{enumerate}

The probability of each outcome can be multiplied by $\mu_l$ or $\mubar_l$ to give the probability that it will or will not be received by $\rx_l$, respectively.

\blue{Note} that once the sender begins mode $l$ leader transmissions of a new packet, it will continue to do so until \red{it is seen, i.e.} $\rx_l$ receives either a mode $l$ or \red{mode $1,\cdots,l-1$} knowledge differential transmission \blue{of that packet}. Therefore, $L_l^*(k)$ can be considered as the probability there have been $k$ unreceived mode $l$ transmissions since the most recently received mode $l$ or knowledge differential transmission.

\red{In} this context, mode $1,\cdots,l-1$ non-knowledge differential transmissions as well as mode \blue{$l+1,\cdots,R$} transmissions are not of interest, since they have no impact on the number of mode $l$ leader transmissions. \red{These correspond to items 3 and 4 in our list of sender actions. We can ignore these transmissions by normalising the probability of each event of interest by $\beta_l+\mu_l\gamma_l$, the likelihood that item 1 or 2 from our list will occur.}

\red{Then the} probability that an unreceived mode $l$ transmission occurs \red{before either a knowledge differential transmission or a received mode $l$ transmission} is given by
 \beq \frac{\mubar_l \beta_l}{\beta_l+\mu_l\gamma_l} \label{eq:unrxfirst} \eeq
and the probability of a received mode $l$ or knowledge differential transmission occurring \red{instead of an unreceived mode $l$ transmission} is given by
 \beq \frac{\mu_l\beta_l+\mu_l\gamma_l}{\beta_l+\mu_l\gamma_l}. \label{eq:anyrx}\eeq
We are now ready to determine the value of $L^*_l(k)$. For there to have been exactly $k$ prior mode $l$ leader transmissions of $\rx_l$'s current next packet $\nnt{l}$, two requirements must be satisfied. Firstly, for there have been at least $k$ transmissions of $\nnt{l}$, the previous $k$ transmissions of interest must have been unreceived mode $l$ transmissions. Secondly, to ensure that there have been no more than $k$ transmissions of $\nnt{l}$, these unreceived mode $l$ transmissions must have been preceded by either a received mode $l$ or knowledge differential transmission. Combining \eqref{eq:unrxfirst} and \eqref{eq:anyrx}, $L_l^*(k)$ is then given by
 \beq L_l^{*}(k)=\left(\frac{\mubar_l \beta_l}{\beta_l+\mu_l\gamma_l}\right)^{k}\frac{\mu_l\beta_l+\mu_l\gamma_l}{\beta_l+\mu_l\gamma_l}.\label{eq:Lstar}\eeq

\subsection{Calculating $L_l(k)$} \label{subsec:Lcalc}

In this part, we calculate the probability $L_l(k)$ that a packet $\pkt_n$ that has been delivered by $\rx_l$ was transmitted $k$ times in total \red{in mode $l$.} The calculation of $L_l(k)$ depends on two separate probabilities: $U_l$, the probability that $\pkt_n$ was unseen by $\rx_l$ from mode $1,\cdots, l-1$ \red{leader transmissions of $\pkt_n$}, and $T_l(k)$, the probability that, given $\pkt_n$ was unseen $\rx_l$ from mode $1,\cdots,l-1$ transmissions, there were exactly $k$ mode $l$ leader transmissions of $\pkt_n$. We now calculate these values.


\subsubsection{Calculating $U_l$}

We begin by calculating $U_l$, the probability that a packet $\pkt_n$ will remain unseen to receiver $\rx_l$ from mode $1,\cdots,l-1$ \red{leader transmissions of the packet}.
The probability that $\rx_l$ will not see $\pkt_n$ from mode $i$ transmissions, where $i<l$, is given by
\beq \sum_{k=0}^{\infty}L_i(k)\mubar_l^k. \label{eq:Lunseen}\eeq
Under Approximation \ref{ap:indep}, we can assume that seen packets are distributed uniformly across the buffer. Therefore the probability that a packet $\pkt_n$ will still be unseen by \blue{$\rx_l$} from mode \blue{$1,\cdots,l-1$} transmissions is given by
 \beq U_l=\prod_{i=1}^{l-1} \sum_{k=0}^{\infty}L_i(k)\mubar_l^k. \label{eq:mprunseen}\eeq

\subsubsection{Calculating $T_l(k)$}

Next we find $T_l(k)$, the probability that \red{there will be exactly $k$ mode $l$ leader transmissions of a packet $\pkt_n$, given it was unseen from mode $1,...,l-1$ leader transmissions. As in Section \ref{subsec:Lstcalc}, only mode $l$ transmissions and received knowledge differential transmissions impact the value of $k$. So again, we ignore all other transmissions by normalising the probability of each relevant transmission by $\beta_l+\gamma_l\mu_l$.}

If $k=0$, then a knowledge differential transmission must have been received \red{before any mode $l$ leader transmissions of $\pkt_n$ could occur}. This occurs with probability
 \beq T_l(0) = \frac{\mu_l\gamma_l}{\beta_l+\mu_l\gamma_l}. \label{eq:L1}\eeq

For $k>0$, using \eqref{eq:unrxfirst} we find that the probability of $k-1$ unreceived mode $l$ transmissions occurring prior to any receptions of knowledge differential or mode $l$ transmissions is given by
 \beq \left( \frac{\mubar_l \beta_l}{\beta_l+\mu_l\gamma_l} \right)^{k-1}. \label{eq:kunrx} \eeq

\red{Following these $k-1$ unreceived mode $l$ transmissions,} one of two things can happen \red{to limit the sender to exactly $k$ mode $l$ leader transmissions of $\pkt_n$.}
\begin{itemize}
 \item A received mode $l$ transmission occurs,
 \item An unreceived mode $l$ transmission occurs, followed by a received knowledge differential transmission.
\end{itemize}
\red{The probability that either of two events will occur is given by}
 \beq \frac{\mu_l \beta_l}{\beta_l+\mu_l\gamma_l}+ \frac{\mubar_l \beta_l\mu_l\gamma_l}{(\beta_l+\mu_l\gamma_l)^2}. \label{eq:endunrx}\eeq
Combining \eqref{eq:L1}, \eqref{eq:kunrx} and \eqref{eq:endunrx}, the probability that exactly $k$ mode $l$ transmissions are required to obtain a previously unseen packet $\nnt{l}$ is given by
 \beq T_l(k)=\begin{cases}
 \frac{\mu_l\gamma_l}{\beta_l+\mu_l\gamma_l}, & k=0\\
 \frac{\mu_l\beta_l(\beta_l+\gamma_l)(\mubar_l \beta_l)^{k-1}}{(\beta_l+\mu_l\gamma_l)^{k+1}}, & k>0.
 \end{cases}\label{eq:T}\eeq


We now scale this by $U_l(k)$, which from \eqref{eq:mprunseen} is the probability that $\pkt_n$ is unseen after mode $1,\cdots,l-1$ transmissions, to obtain the probability that the sender will transmit a packet $k$ times,
 \beq L_l(k)=\begin{cases}
  1-U_l+U_lT_l(0),& k=0\\
  T_l(k) U_l, & k>0.
 \end{cases} \label{eq:mLak}\eeq

\subsection{Comparison with simulation}

\begin{figure}\begin{center}
\includegraphics[width=\picsize]{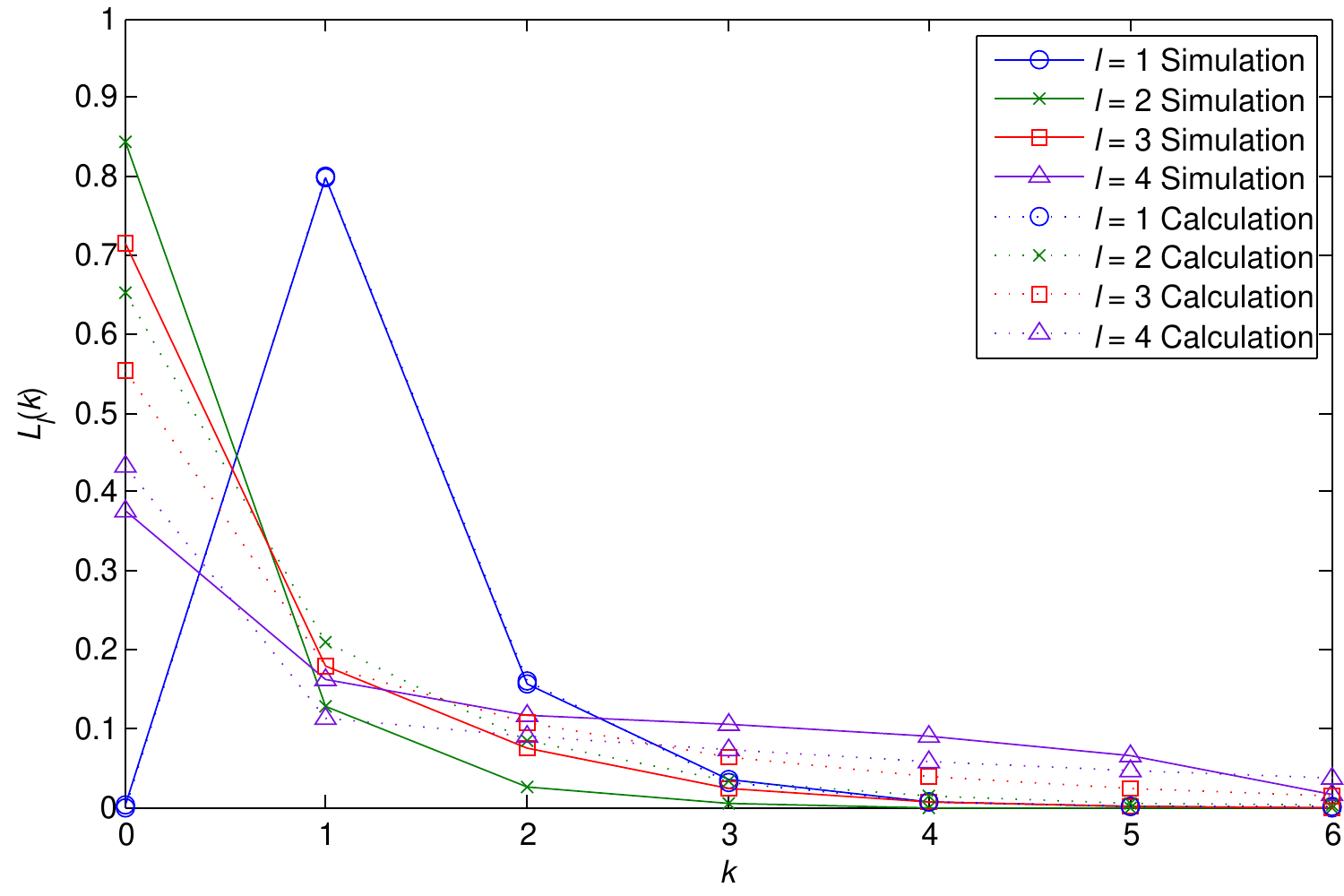}
\caption{Probability of $k$ leader transmissions of a packet, for $\muv=[0.8,0.6,0.4,0.2]$ and $\betav=[0.85,0.05,0.05,0.05]$.}\label{fig:mu822betaL}
\end{center}\end{figure}

We examine the accuracy of our $L_l(k)$ calculations by comparing the calculated and simulation-based probabilities in Fig. \ref{fig:mu822betaL}. We observe that calculated and simulation values match extremely well for receivers $l=1,2,3$, but loses some accuracy for $\rx_4$, although it does follow the same general trend. This can be attributed to the fact our \med{approximations lead to} some loss of accuracy, particularly for higher receiver indices.

We now examine the shape of $L_l(k)$, for different values of $l$. For $l=1$, every packet must be transmitted at least once for $\rx_1$ to receive it. Therefore there is no possibilility of $k=0$ for this receiver. The most likely case is $k=1$, since $\rx_1$'s high channel rate means the majority of the time, a packet will be received after one transmission. For the same reason, $L_1(k)$ decays relatively quickly as $k$ increases. At higher values of $l$, however, it is reasonably likely that $k=0$ because the receiver can overhear the packet from faster receivers' leader transmissions. Although the number of leader transmissions a receiver can overhear increases as the receiver index increases, the receiver's channel rate also decreases. In this simulation, it results in a smaller probability of $k=0$ as $l$ increases. We also observe that the distribution of $L_l(k)$ decays more slowly for higher values of $l$. This can also be attributed to the lower channel rates of the corresponding receivers.

%% file: kdiff_mkdiff.tex
\section{Knowledge differentials}\label{supsec:mkdiff}

In this section, we will determine $K_r^m$, the probability that a receiver $\rx_r$ obtains a knowledge differential from a mode $m$ transmission, given the mode vector $\betav$. Through the dependencies on $L_l^*(k)$ and $L_l(k)$, all calculations in this section are also implicitly dependent on $\betav$, but for readability we omit direct mention of $\betav$ in our notation.

\subsection{Knowledge differences}\label{subsec:mkdiffc}

We calculate $D_a^b$, the probability of a receiver $\rx_a$ having a $\rx_b$ knowledge difference. Recall from Definition \ref{def:kdiffc} that if $a\leq b$ then $D_a^b=1$.

If  $b<a$, then the probability $D_a^b$ that a receiver $\rx_a$ has a $\rx_b$ knowledge difference is given by the probability that $\rx_a$ has seen $\nnt{b}$ from at least one of the \med{ mode $1,\cdots,b$ transmissions. This can be determined from the \pmed{complementary} probability of the packet \emph{not} being seen from mode $1,\cdots,b$ transmissions, which can be calculated using \eqref{eq:Lunseen} and \eqref{eq:mprunseen}.}
\beq D_a^b=1-\left(\sum_{k=0}^{\infty}L_b^{*}(k)\mubar_a^k\right)\prod_{l=1}^{b-1}\sum_{k=0}^{\infty}L_l(k)\mubar_a^k. \label{eq:kdiffc}
\eeq
\blue{Recall that in mode $b$ transmissions, $\nnt{b}$ is the next needed packet of $\rx_b$ and hence $L_b^{*}(k)$ is used instead of $L_l(k)$.}

\subsection{Probability of a coding}
We next establish the probability $P^m(C)$ that in mode $m$, a particular coding $C$ will occur. The notation $P^m(\cdot)$ represents the mode $m$ probability of the event inside the brackets.

\begin{definition}
The \emph{subcoding} $C_r$ takes the first $r$ elements of the coding $C$. $C_r$ is the $r$-length binary vector representing which of receivers $\rx_1$ to $\rx_{r}$'s next needed packets have been coded into $s(t)$.
\begin{example}
For the coding $C=[1, 0, 1, 1]$, the corresponding subcodings are $C_1=[1]$, $C_2=[1, 0]$, $C_3=[1, 0, 1]$ and $C_4=[1,0,1,1]$.
\end{example}
\end{definition}

Combining Property \ref{pr:txcod}, Property \ref{pr:kdiff} and Approximation \ref{ap:uncod}, a packet $\nnt{r}$ will only be coded into $s(t)$ if $\rx_r$ \red{has seen all of the} other packets coded into $s(t)$. But by Approximation \ref{ap:orderednnt}, receiver $\rx_r$ has already decoded the next needed packets of receivers $\rx_{r+1},\cdots,\rx_R$. Therefore the subcoding $C_r$ corresponds to the portion of the transmission coding $C$ that will affect whether or not $\nnt{r}$ is coded into $s(t)$, and therefore the value of $C(r)$. From this information, we can break down finding the probability $P^m(C)$ into the problem of finding the probability , $P^m(C_r)$ for $r=1,\cdots,R$.

In practice, it is usually not necessary to perform calculations starting from $P^m(C_1)$. A mode $m$ transmission will always code the next needed packet of the leader, so
 \beq P^m(C(m)=1)=1. \label{eq:Cleader}\eeq
Furthermore, since $\nnt{1},\cdots,\nnt{m-1}$ will not be coded into a mode $m$ transmission,
 \beq C(1)=\cdots=C(m-1)=0. \label{eq:Cuncod} \eeq
Combining these two pieces of information, we obtain the result,
 \beq P^m(C_m)=\begin{cases}
  1, & \mbox{for } C_m=[0,\cdots,0,1]\\
  0, & \mbox{for all other } C_m.
 \end{cases}\label{eq:PrC1}
 \eeq
This means that we can start our calculations from $C_m$, since all possible mode $m$ subcodings begin with $[0,\cdots,0,1]$. Equation \eqref{eq:PrC1} can be used as the base case to determine the probability of the other subcodings $C_r$, for $r=m+1,\cdots,R$.

Under Property \ref{pr:kdiff} the only way for a packet $\nnt{r}$ to be coded is if the corresponding receiver $\rx_r$ has seen all packets coded into the transmission $s(t)$. Then the probability of $\nnt{r}$ being added to $s(t)$ is the probability that $\rx_r$ has a $\rx_b$ knowledge difference for all receivers $\rx_b$ corresponding to packets coded into $s(t)$,
  \beqa P^m(C(r)=1|C_{r-1})&=&\smashoperator[r]{\prod_{b\in f_1^m(C_{r-1})}}\: D_r^b\nonumber\\
  &=& \red{\prod_{b\in f_1^m(C)} D_r^b}, \label{eq:kdiffc2kdiffl}\eeqa
where for a vector $v$, we define $f_1^m(v)$ as the list of all indices $r\geq m$ where $v(r)=1$, and $f_0^m(v)$ as the list of all indices $r>m$ where $v(r)=0$. \red{Note that the second part of this equation is true because $D_r^b=1$ for all $r\leq b$.}

The probability $P^m(C_r)$ of each subcoding $C_r$ can then be defined in terms of the previously calculated probability $P^m(C_{r-1})$,
 \beq P^m(C(r)=1)=P^m(C_{r-1})P^m(C(r)=1|C_{r-1}),\eeq
 \beq P^m(C(r)=0)=P^m(C_{r-1})P^m(C(r)=0|C_{r-1}). \eeq
Combining this pair of equations with \eqref{eq:kdiffc2kdiffl}, we \red{can derive} the probability of a particular subcoding $C_r$:
\red{
 \beq P^m(C_r)=\left( \prod_{a\in f_1^m(C_r)} \smashoperator[r]{\prod_{b \in f_1^m(C_{r})}}  D_a^b\right) \left(\prod_{a\in f_0^m(C_r)}1-\smashoperator{\prod_{b\in f_1^m(C_{r})}} D_a^b\right), \label{eq:CfromD} \eeq
 }
if the first $m$ elements of $C_r$ are $C_m=[0,\cdots,0,1]$. Otherwise, $P^m(C_r)=0$.

\subsection{Knowledge differentials} \label{subsec:mkdiff}

We now find the probability $K_r^m$ of a knowledge differential occurring for a receiver $\rx_r$ under  \blue{mode $m$ transmission}.

By Algorithm \ref{alg:simplecod} \blue{receiver $\rx_r$'s} next needed packet is coded iff a knowledge differential occurs. This means that \red{we can determine the probability of a knowledge differential by finding the probability that the next needed packet of $\rx_r$ is coded into a transmission,}
 \beqa K_r^m&=&P^m(C(r)=1)\nonumber\\
 &=& \red{\sum_{C}  C(r) P^m(C)} \nonumber\\
 &=&\sum_{C_r} C_r(r) P^m(C_r).\label{eq:mkdiff}\eeqa

It can be noted that, in the special case where $r=m$, we can substitute \eqref{eq:PrC1} to obtain
\beq K_m^m=1. \label{eq:leadkdiff}\eeq

\begin{property}
The probability $K_r^m$ of a mode $m$ knowledge differential for a receiver $\rx_r$ depends only on the ratio of
$\beta_1:\beta_2:\cdots:\beta_r$. \label{pr:mbuffer}
\begin{IEEEproof}
\upshape{We use induction, starting with the base case of $r=1$. From \eqref{eq:leadkdiff} and Property \ref{pr:minnov}, $K_1^1=1$, and $K_1^m=0$ for $m>1$. This is true regardless of what is happening to the other receivers, so for \blue{the fastest receiver corresponding to $r=1$,} Property \ref{pr:mbuffer} is satisfied.

For each subsequent value of $r$, we examine the dependency of $K_r^m$ on $\betav$ using Fig. \ref{fig:blockanalysis}. We observe that dependencies on the \blue{mode vector $\betav$} are fed in through two variables: $T_l(k)$ and $L^*_l(k)$, via $\gamma_l$. In particular, $K_r^m$ is calculated from $T_1(k),\cdots,T_{r-1}(k)$ and $L^*_r(k)$. Therefore we need only prove that $T_l(k)$ and $L^*_l(k)$ depend on the ratio $\beta_1:\beta_2:\cdots:\beta_{l}$.

From \eqref{eq:gamma}, if $K_l^i$ satisfies Property \ref{pr:mbuffer} for $i<l$, then scaling $\beta_1,\cdots,\beta_l$ by some factor $\scaling>0$ will also scale $\gamma_l$ by a factor of $\scaling$. We then make the key observation that, in each case of \eqref{eq:T}, \emph{every term in the numerator and denominator has the same multiplicity of either $\beta_l$ or $\gamma_l$.} \med{This means that multiplying $\beta_1,\cdots,\beta_l$ and $\gamma_l$ by a factor of $\scaling$ in \eqref{eq:T} will not change the value of $T_l(k)$.} Therefore, scaling $\beta_1,\cdots,\beta_{l}$ by some value $\scaling$ will not affect the value of $L_{l}(k)$.

In \eqref{eq:Lstar}, we similarly observe that the same multiplicity of $\beta_l$ and $\gamma_l$ in the numerator and denominator of $L^*_l(k)$. Therefore both $T_l(k)$ and $L_l^*(k)$ depend only on the ratio $\beta_1:\beta_2:\cdots:\beta_{l}$.}
\end{IEEEproof}
\end{property}

\subsection{Simulation results}

We examine the accuracy of our knowledge differential calculations. We can observe here that, for $r=1,2,3$ the calculated values of $K_r^m$ match the simulated values extremely well, while at $r=4$ some inaccuracy observed at $m=1,2$. This can be attributed to the cross-receiver effects from $\rx_2$ and $\rx_3$ in modes $m=1,2$. For $m=3,4$ only $\rx_3$ and $\rx_4$'s packets are coded. Since only $\rx_3$ can affect $\rx_4$'s knowledge differential probability, there are few cross-receiver interactions to take inot account when determining $K_r^m$, so the calculations remain accurate.

The shape of the $K_r^m$ can be explained as follows. For a given receiver $\rx_r$, $K_r^m$ is a relatively low probability at $m=1$, that increases as $m$ increases. This increasing probability can be attributed to the lesser number of coded receivers, which makes it more likely that a knowledge differential will occur. At mode $m=r$, $K_r^m$ reaches its maximum value, with $K_r^r=1$. This matches with \eqref{eq:leadkdiff}. In modes $m>r$, transmissions are no longer innovative, in line with  Property \ref{pr:minnov}. As a result, the probability of a knowledge difference is $K_r^m=0$.

\begin{figure}\begin{center}
\includegraphics[width=\picsize]{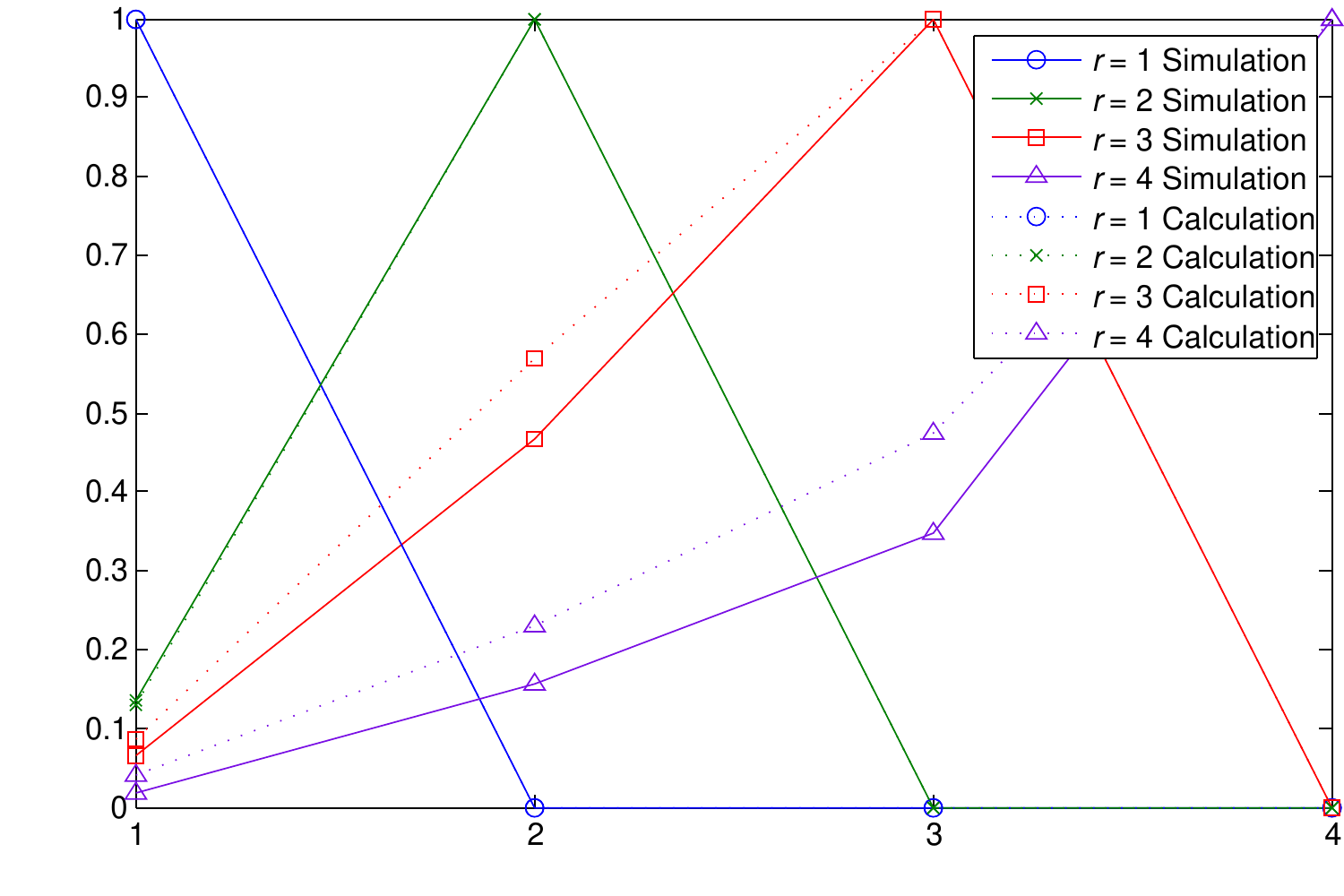}
\caption{Probability of a knowledge differential for a receiver $\rx_r$ in mode $m$, for $\muv=[0.8,0.6,0.4,0.2]$ and $\betav=[0.85,0.05,0.05,0.05]$.}\label{fig:mu822betaK}
\end{center}\end{figure}

%% file: kdiff_mrates.tex
\section{Delivery Rates}\label{sec:mrates}

Now that the probability of $K_r^m$ of a mode $m$ knowledge differential is known, we are ready to estimate $R_r$, the delivery rate of each receiver $\rx_r$. It can be observed in Fig. \ref{fig:blockanalysis} that $B_r$ and $R_r$ are recursively dependent on one another.

As outlined in Section \ref{sec:mod:buffer}, we can divide all transmissions received by $\rx_r$ into three categories. Those that are non-innovative,
those that allow the delivery of $\nnt{r}$, and those that contribute to the buffer density. We use this information to estimate the delivery rate for the receivers.

\subsection{Buffer density} \label{sec:buffer}

We firstly determine $B_r$, receiver $\rx_r$'s buffer density. The first step is to determine which of the sender's transmissions can contribute to the buffer density. \red{From Section \ref{subsec:bufregions}, a transmission must be both innovative and a non-knowledge differential transmission in order to be stored in the packet density region of the buffer.} By Property \ref{pr:minnov}, transmissions from modes $m=r+1,\cdots,R$ will be noninnovative to $\rx_r$. From Section \ref{supsec:mkdiff}, $K_r^m$ is the fraction of the sender's mode $m$ transmissions that will result in knowledge differentials for $\rx_r$ (for $m\leq r$), \red{and from \eqref{eq:leadkdiff} $K_r^r=1$. Therefore only the $\overline{K_r^m}$ fraction of mode $m=1,\cdots,r-1$ non-knowledge differential transmissions can contribute to $\rx_r$'s buffer density.}

Since the distribution of leader transmissions studied in Section \ref{sec:Lcalc} is uniform, it is reasonable to say that within a subregion, the buffer density will also be uniform. Therefore, the buffer density $B_r$ is the average number of non-knowledge differential transmissions \red{received from modes $m=1,\cdots,r-1$ at each packet index in subregion $r$}.

\red{We consider the contribution from each mode $m<r$. For a given mode $m$, we assume all received mode $m$ non-knowledge differential packets are distributed uniformly between $\pkt_1,\cdots,\nnt{m}$. Then the average number of sender transmissions corresponding to each packet index in subregions $m,\cdots,R$ is the inverse of $\rx_m$'s delivery rate, $1/R_m$.} Of the $\beta_m$ fraction of time spent in mode $m$, $\overline{K_r^m}$ of the packets transmitted will be non-knowledge differential transmissions \red{for $\rx_r$}. These are distributed uniformly across $\pkt_1,\cdots,\nnt{m}$ so there is on average $\beta_m/R_m$ mode $m$ transmissions per packet. Of these, $\mu_r$ will be received by $\rx_r$. So summing over all modes $m=1,\cdots,r-1$,
 \beq B_r=\mu_r \sum_{m=1}^{r-1} \frac{\beta_m\overline{K_r^m}}{R_m}. \label{eq:mbuffer} \eeq
For $r=1$ there would be no summation, so we use this as the base case for our recursive calculation,
 \beq B_1=0. \label{eq:buf1} \eeq

\subsection{Delivery rate}\label{subsec:rateans}
The task of delivering new packets to a receiver $\rx_r$ is equivalently the process of making the unseen packets in the buffer seen to $\rx_r$, using knowledge differential transmissions. On average, $B_r$ of the undelivered packets in the buffer have already been seen by the receiver. Therefore it is
necessary to fill in the remaining $(1-B_r)$ fraction of packets using knowledge differential transmissions. On average, \blue{$\mu_r$} of the knowledge differential transmissions will be received. Therefore, the delivery rate is given by
\blue{ \beq R_r=\frac{\mu_r\sum_{m=1}^r \beta_m K_r^m}{1-B_r}. \label{eq:mrate} \eeq}
Here we must sum the knowledge differentials contributed by each mode. Note that the summation only goes from 1 to $r$, since from Property \ref{pr:minnov}, a mode $m$ transmission can only be innovative to receivers $\rx_m,\cdots,\rx_R$.

We can observe from \eqref{eq:mbuffer} and \eqref{eq:mrate} that there is a recursive relationship between the delivery rates of the receivers. Note that $B_r$ only depends on $R_1,\cdots,R_{r-1}$ and $K_r^m$, calculated in Section \ref{supsec:mkdiff}, while $R_r$ only depends on $B_r$ and $K_r^m$. Therefore starting from the $\rx_1$ base case, where the delivery rate is $R_1^m=\beta_1 \mu_1$ and the buffer density is $B_1^m=0$, we can sequentially calculate the delivery rates of all receivers.

\subsection{Rate calculation algorithm}
The process for calculating the delivery rate has many recursive dependencies that make it difficult to determine how to implement. Therefore, we provide Algorithm \ref{alg:ratecalc} to outline the order in which variables shown in Fig. \ref{fig:blockanalysis} can be calculated.

Making use of the fact that a receiver's delivery rate is only influenced by \blue{faster} receivers, we sequentially calculate the delivery rate $R_r$ \blue{for} each receiver $\rx_r$ starting from $r=1,\cdots,R$. Following the same structure as Sections \ref{sec:Lcalc} to \ref{subsec:rateans}, we perform the leader transmission \red{calculations}, followed by knowledge differential calculations and finally the delivery rate calculations.

It is interesting to note that, in the special case of $r=1$, the \red{calculations} reduce significantly. Because there are no earlier receivers for $\rx_1$'s behaviour to be influenced by, the calculation dependencies illustrated by curved arrows in Fig. \ref{fig:blockanalysis} can be removed.\footnote{\red{In our calculations,} empty summations are equal to zero, and empty products are equal to one.}

\begin{algorithm}
\caption{Delivery rate calculation.}\label{alg:ratecalc}
\begin{algorithmic}[1]
\STATE Set $D_a^b=1$ for all $a\geq b$.
\STATE Set $K_r^m=1$ for all $r=m$, and $K_r^m=0$ for $r<m$.
\FOR {$m=1,\cdots,R$}
  \STATE Calculate $\gamma_m$, using $K_m^1,\cdots,K_m^{m-1}$.
	\STATE Calculate $L_m^*(k)$, using $\gamma_m$.
	\STATE Calculate $U_m$ using $L_1,\cdots,L_{m-1}(k)$.
  \STATE Calculate $T_m(k)$ using $\gamma_m$
  \STATE Calculate $L_m(k)$ using $U_m$ and $T_m(k)$.
	\FOR {$a=m+1,\cdots,R$}
	  \STATE Calculate $D_a^m$ using $L_1(k),\cdots,L_{m-1}$ and $L_m^*(k)$.
	\ENDFOR
	\FOR {$n=1,\cdots,m$}
	  \FOR {all $C_m$}
	    \STATE Calculate $P^n(C_m)$ using $D_a^b$ for $b<a<m$.
	  \ENDFOR
	  \FOR {$r=n,\cdots,m$}
	    \STATE Calculate $K_r^n$ using all $P^n(C_r)$.
    \ENDFOR
	\ENDFOR
  \STATE Calculate $B_m$ using \blue{$K_m^1,\cdots,K_m^{m-1}$} and $R_1,\cdots,R_{m-1}$.
  \STATE Calculate $R_m$ using $K_m^1,\cdots,K_m^m$ and $B_m$.
\ENDFOR
\end{algorithmic}
\end{algorithm}

\subsection{Simulation results}

\begin{figure}\begin{center}
\includegraphics[width=\picsize]{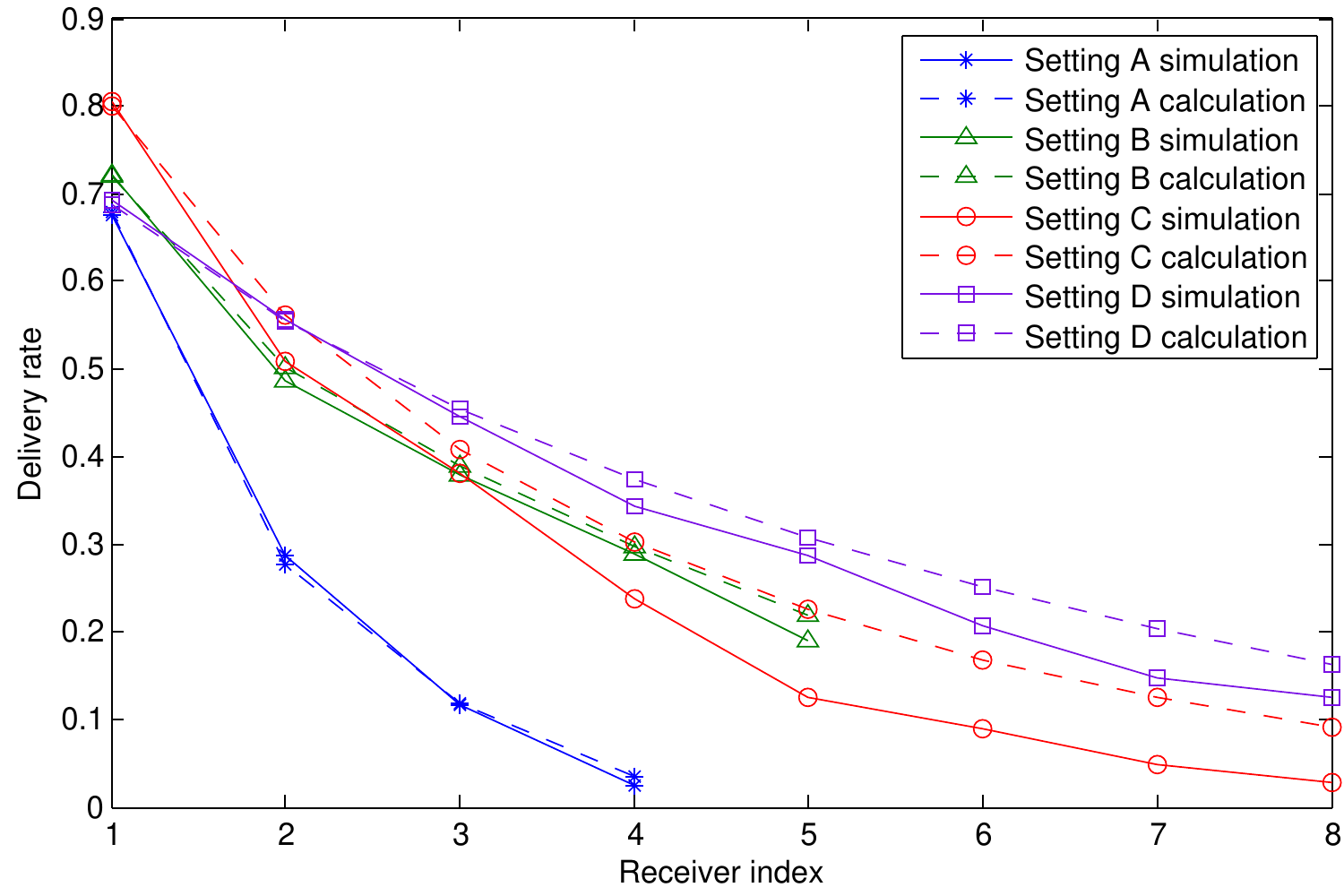}
\caption{Probability of a knowledge differential for a receiver $\rx_r$ in mode $m$, for the simulation settings given in Table \ref{tab:simconditions}.}\label{fig:rates}
\end{center}\end{figure}

We now compare the simulated and calculated delivery rates for the receivers. Fig. \ref{fig:rates} illustrates the simulated and calculated delivery rates of a set of receivers under the settings listed in Table \ref{tab:simconditions}.

Under Settings A and B, the calculated delivery rates match well with the values observed in simulation. In settings C and D on the other hand\med{, the approximate calculations} are fairly accurate for the first few receivers, but gradually lose accuracy for higher receiver indices. The calculated values tend to be a little optimistic as the receiver index grows, since at higher receiver indices, Approximation \ref{ap:uncod} loses its accuracy. Setting C represents the extreme case, where all sender transmissions are allocated to mode 1. For a receiver $\rx_r$, the closer the mode $m$ is to $r$, the more accurate the knowledge differential estimate will be. Therefore greater the amount of time allocated to higher modes, the more accurate the calculated rates will be.\footnote{Note, however, that $\betav$ must be restricted to values that maintain the ordering of the receivers under Approximation \ref{ap:orderednnt}.}

It is interesting to note that, for Setting A, despite some slight inaccuracies in the knowledge differential estimates of Fig. \ref{fig:mu822betaK}, the delivery rates under both simulation and calculation match quite well. In part, this can be attributed to the inverse relationship between the buffer density and the probability of knowledge differential that can be observed in \eqref{eq:mbuffer}. As a result of this relationship, inaccuracies in the calculated probability of a knowledge differential are partially mitigated by their inverse effect on the buffer density.

\begin{table}\begin{center}
\begin{tabular}{|c|c|c|}
\hline Setting  & $\muv$ & $\betav$\\
\hline A & $[0.8,0.6,0.4,0.2]$ & $[0.85,0.05,0.05,0.05]$\\
\hline B & $[0.9,0.8,0.7,0.6,0.5]$ & $[0.8, 0.05,\cdots,0.05]$\\
\hline C & $[0.8,0.75,0.7,\cdots,0.45]$ & $[1,0,\cdots,0]$\\
\hline D & $[0.8,0.75,\cdots,0.45]$ & $[0.86,0.02,\cdots,0.02]$\\
\hline
\end{tabular}
\tablespace
\caption{Simulation conditions used in Fig. \ref{fig:rates}.}\label{tab:simconditions}
\end{center}\end{table}

%% file: kdiff_fairness.tex
\section{Fairness} \label{sec:fair}
We now use the delivery rate analysis of Sections \ref{supsec:mkdiff} and \ref{sec:mrates} to determine the value of the mode vector $\betav$ that will result in fair delivery rates for all receivers. We define fairness as follows:
\begin{definition}
A transmission scheme is \emph{fair} for all receivers if
their delivery ratios are equal, so that
 \beq \frac{R_1}{\mu_1}=\frac{R_2}{\mu_2}= \cdots=\frac{R_R}{\mu_R}.
 \eeq
\end{definition}

Our aim then is to achieve equal \emph{delivery ratios} for all receivers, where the delivery ratio $Q_r$ of a receiver $\rx_r$ is
defined as
 \beq Q_r=\frac{R_r}{\mu_r} \label{eq:Q}\eeq

In practice, the method we will outline can be modified to equalise any set of delivery ratios, so long as doing so will maintain the condition
 \beq R_1\geq R_2\geq \cdots\geq R_R. \eeq

\subsection{\red{The rate scaling property}}
Before we can begin, it is necessary to highlight the key property of our transmission scheme that
will help us build our fairness algorithm.

\begin{property}\label{pr:ratescaling}
Scaling $\beta_1,\cdots,\beta_r$ by some amount $\scaling>0$ does not affect the buffer density of receivers $\rx_1,\cdots,\rx_r$, and scales their delivery rates by $\scaling$.

\begin{IEEEproof}\upshape{
We will prove this property using induction. Let us substitute $\beta'_m=\scaling \beta_m$ for all $m=1,\cdots,r$ into \eqref{eq:mrate} and \eqref{eq:mbuffer}. Using the notation $B_r(\beta)$ and $R_r(\beta)$ to show the explicit dependency on the mode vector $\betav$, we obtain
\beq B_r(\beta')=\mu_r \sum_{m=1}^{r-1} \frac{\scaling \beta_m\overline{K_r^m}} {R_m(\beta')} \label{eq:modebufferP} \eeq
\beq R_r(\beta')=\frac{\sum_{m=1}^r \scaling \beta_m K_r^m}{1-B_r(\beta')} \label{eq:moderateP} \eeq
As the base case, $R_1(\beta')=\scaling \beta_1 \mu_1$ and $B_1(\beta')=0$. $R_1(\beta')$ is linearly dependent on $\scaling$, and $B_1(\beta')$ is a constant. Therefore Property \ref{pr:ratescaling} holds for the base case.

We now show that, if Property \ref{pr:ratescaling} is true for $\rx_1,\cdots,\rx_{r-1}$, then it also holds for $\rx_{r}$. From Property \ref{pr:mbuffer}, we know that scaling $\beta_1,\cdots,\beta_r$ by $\scaling$ will not affect the probability $K_r^m$ of a knowledge differential. Now, assume that $R_m(\beta')=\scaling R_m(\beta)$ and $B_m(\beta')=B_m(\beta)$ for $m<r$. Then by substituting these values into \eqref{eq:moderateP} and \eqref{eq:modebufferP}, we obtain

 \beqa B_r(\beta')&=&\mu_r \sum_{m=1}^{r-1} \frac{\scaling \beta_m\overline{K_r^m}} {\scaling R_m(\beta)}\nonumber\\
 &=& B_r(\beta) \eeqa
and
 \beqa R_r(\beta')&=&\frac{\sum_{m=1}^r \scaling \beta_m K_m^r}{1-B_r(\beta)}\nonumber\\
&=&\scaling R_r(\beta). \eeqa
Therefore, Property \ref{pr:ratescaling} holds for all receivers.}
\end{IEEEproof}
\end{property}
This is a key result since it tells us that, for a receiver $\rx_r$, scaling the time spent in modes $1,\cdots,r$ by a factor of $\scaling$ will scale the delivery \red{rates $R_1,\cdots,R_r$} by a factor of $\scaling$ as well. \red{This will give us a predictable way to determine the receivers' delivery rates as we adjust the value of the mode vector $\betav$.}

\subsection{Fairness algorithm}

We now outline the fairness algorithm used to determine the value of $\betav$ required to achieve delivery rate fairness. Recall that, in Section \ref{sec:mrates}, we calculated both the expected buffer density and delivery rates, given $\betav$. The aim now is to determine the value of $\betav$ that will result in fair delivery rates for all receivers.

The receivers' delivery ratios can be matched using the procedure outlined in Algorithm \ref{alg:fairness}. \med{This fairness algorithm performs two functions as it iterates through each value of $r$. Firstly, it scales down the amount of time allocated to modes $1,\cdots,r-1$ by the calculated quantity $1-\scaling[r]$. Secondly, it allocates the remaining $\scaling[r]$ fraction of time to mode $r$. By Properties \ref{pr:minnov} and \ref{pr:ratescaling}, neither of these actions affects the delivery ratio of receivers $1,\cdots,r-1$. Therefore, Algorithm \ref{alg:fairness}} ensures that equalised delivery ratios will remain matched through subsequent iterations of the algorithm.  After the $r\ith$ iteration is complete, $\betav[r]$ will allow \red{the sender to achieve} $Q_1[r]=\cdots= Q_r[r]$.

\med{\subsubsection{Notation}
To reflect the changing values at each iteration, we append the variables with an index $[r]$ to represent the updated values as of the $r^{\mbox{th}}$ iteration. $\beta_m[r]$ represents the time allocated to mode $m$ after iteration $r$, and $\scaling[r]$ is an} equalisation factor which determines $\betav[r]$ based on $\betav[r-1]$, the value at the previous iteration. The delivery ratio $Q_m[r]$ of each receiver $\rx_r$ is also determined from $\betav[r]$. At each iteration $r$, the value $\betav[r]$ is updated to match the delivery ratio of $\rx_r$ to that of receivers $\rx_1,\cdots,\rx_{r-1}$. \red{In the remainder of this section, we will explicitly show any dependencies on $\betav[r]$, for example $K_r^m(\betav[r])$ represents receiver $\rx_r$'s mode $m$ knowledge differential probability given that $\betav[r]$ is used as the mode vector.}
\begin{algorithm}
\caption{Fairness algorithm.}\label{alg:fairness}
\begin{algorithmic}[1]
\STATE Set $\beta_1[1]=1$, and $\beta_2[1],\cdots,\beta_R[1]=0$.
\FOR{$r=2$ to $R$}
  \STATE Calculate $Q_{r-1}[r-1]$ and $Q_r[r-1]$ using \eqref{eq:Q}, substituting $\betav=\betav[r-1]$.
  \STATE Calculate $\scaling[r]$ from \eqref{eq:alpha}.
  \STATE Set $\beta_r[r]=\scaling[r]$
\FOR{$m=1$ to $r-1$}
  \STATE Set $\beta_m[r]=(1-\scaling[r])\beta_m[r-1]$
\ENDFOR
\ENDFOR
\STATE \red{Set $\betav=\betav[r]$.}
\end{algorithmic}
\end{algorithm}

We begin by allocating all transmissions to mode 1, so that $\beta_1[1]=1$, and $\beta_m[1]=0$, for
$m=2,\cdots,R$. By \red{\eqref{eq:leadkdiff}, all mode 1} transmissions are knowledge differential transmissions \red{for $\rx_1$}. Therefore the delivery ratio for $\rx_1$ after iteration 1 is $Q_1[1]=1$.

In each iteration $2 \leq r\leq R$ of the for loop, we find a value of $\betav[r]$ which will achieve $Q_1[r]=\cdots =Q_r[r]$, equalising the \red{delivery ratio} of $\rx_r$ with that of receivers $\rx_1$ to $\rx_{r-1}$. This is determined by reducing the time spent in modes $1,\cdots,r-1$ by a factor of $(1-\scaling[r])$ to allow a fraction of time $\scaling[r]$ to be spent in mode $r$.

By Property \ref{pr:ratescaling}, this has the effect of scaling $Q_1[r],\cdots,Q_{r-1}[r]$ by a factor $1-\scaling[r]$, so that
\beq Q_m[r]=(1-\scaling[r])Q_m[r-1] \label{eq:Qrcalc}\eeq
for all $m<r$.

Combining \eqref{eq:mrate}, \eqref{eq:Q}, Property \ref{pr:ratescaling} and \eqref{eq:leadkdiff}, the delivery ratio of receiver $\rx_r$ in iteration $r$ is given by
 \beqa Q_r[r]&=& \red{\frac{\sum_{m=1}^r \beta_m[r] K_r^m(\betav[r])}{1-B_r[r]}}\nonumber\\
 &=&(1-\scaling[r])Q_r[r-1]+\frac{\scaling[r]}{1-B_r(\betav[r-1])}. \label{eq:Qicalc}\eeqa

Therefore to equalise the delivery ratios of $\rx_r$ and $\rx_{r-1}$ the task is to find the value of $\scaling[r]$ for which
$Q_r[r]=Q_{r-1}[r]$. This can be solved using \eqref{eq:Qrcalc} and \eqref{eq:Qicalc} to give
 \beq \scaling[r]=\frac{Q_{r-1}[r-1]-Q_r[r-1]}
 {Q_r[r-1]-Q_{r-1}[r-1]+\frac{1}{1-B_r(\red{\betav[r-1]})}}.\label{eq:alpha}\eeq
For $m=1,\cdots,r-1$ the new values of $\beta_m[r]$ are now given by
 \beq \beta_m[r]=\begin{cases}
 (1-\scaling[r])\beta_m[r-1], & m<r\\
 \scaling[r], & m=r\\
 0, &m>r
 \end{cases}\eeq
 
This process is repeated for each receiver to equalise the delivery ratio of each subsequent receiver, until all receivers are equal.
After this process is complete, the amount of time that should be spent in each mode $m$ to achieve fairness is given by
 \beq \betav=\betav[R]. \eeq
The delivery ratio of every receiver $\rx_r$ using the mode vector $\betav$ will be equal. \med{Substituting \eqref{eq:leadkdiff} and \eqref{eq:buf1} into \eqref{eq:mrate}, the delivery rate of $\rx_1$ is found to be $\mu_1\beta_1$, so by \eqref{eq:Q}, the delivery ratio of $\rx_1$ is $Q_1=\beta_1[R]$.} Therefore, using the mode vector $\betav$ will give the delivery ratios
 \beq Q_1=Q_2=\cdots =Q_R=\beta_1[R]. \eeq

\subsection{Simulation results}
We can now compare the delivery rates achieved both in theory and in practice. The channel rates for each setting we will investigate is given in Table \ref{tab:simsettings}.

\begin{table} \begin{center}
\caption{Settings used in fairness algorithm simulations.}\label{tab:simsettings}
\begin{tabular}{|l|c|}
\hline Setting & $\muv$ \\
\hline A & $[0.8, 0.7, 0.6, 0.5]$\\
\hline B & $[0.9, 0.8, \cdots, 0.5]$\\
\hline C & $[0.9,0.85,\cdots,0.55]$\\
\hline
\end{tabular}
\end{center}\end{table}

\begin{figure}\begin{center}
\includegraphics[width=\picsize]{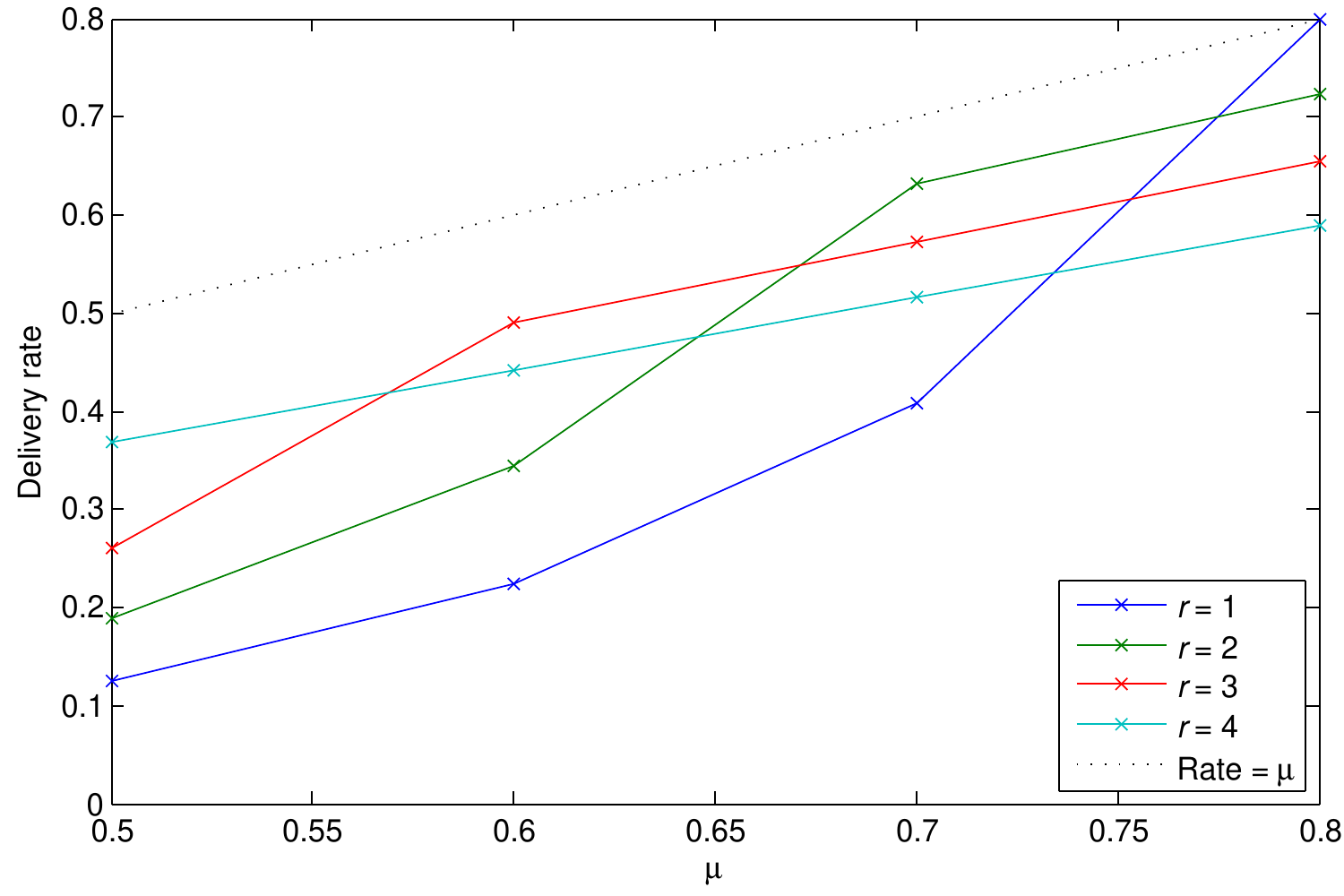}
\caption{Calculated delivery rates achieved after each iteration $r$ of Algorithm \ref{alg:fairness}, under Setting A from Table \ref{tab:simsettings}. These are compared against $\mu$, the channel rate for each receiver.}\label{fig:setAfairness}
\end{center}\end{figure}

In Fig. \ref{fig:setAfairness} we can observe the delivery rate estimates after each iteration $r$ in Algorithm \ref{alg:fairness}. \red{At each iteration $r$, the theoretical delivery rates are those that the sender would enforce if it only took into consideration the delivery ratios} of the first $r$ receivers. \red{At iteration $r=1$}, $\rx_1$ achieves the best possible performance with delivery rate $R_1=\mu_1$, however the other receivers' delivery delays are significantly worse. At each \red{subsequent value of} $r$, the delivery rates of receivers $\rx_1,\cdots,\rx_{r-1}$ are scaled down to allow $\rx_r$ to \red{match the delivery ratios of receivers $\rx_1,...,\rx_{r-1}$.}

\blue{It is interesting to note that, after iteration $r$, there is improvement to the delivery rates of \emph{all} mode $r$ receivers (that is receivers $\rx_r$ to $\rx_R$), not simply $\rx_r$.} This occurs because in higher modes, the transmission $s(t)$ will be coded from \red{a smaller set of receivers, whose next needed} packets are more likely to have been seen. This improves the likelihood that a receiver will obtain a knowledge differential.

\begin{figure}\begin{center}
\includegraphics[width=\picsize]{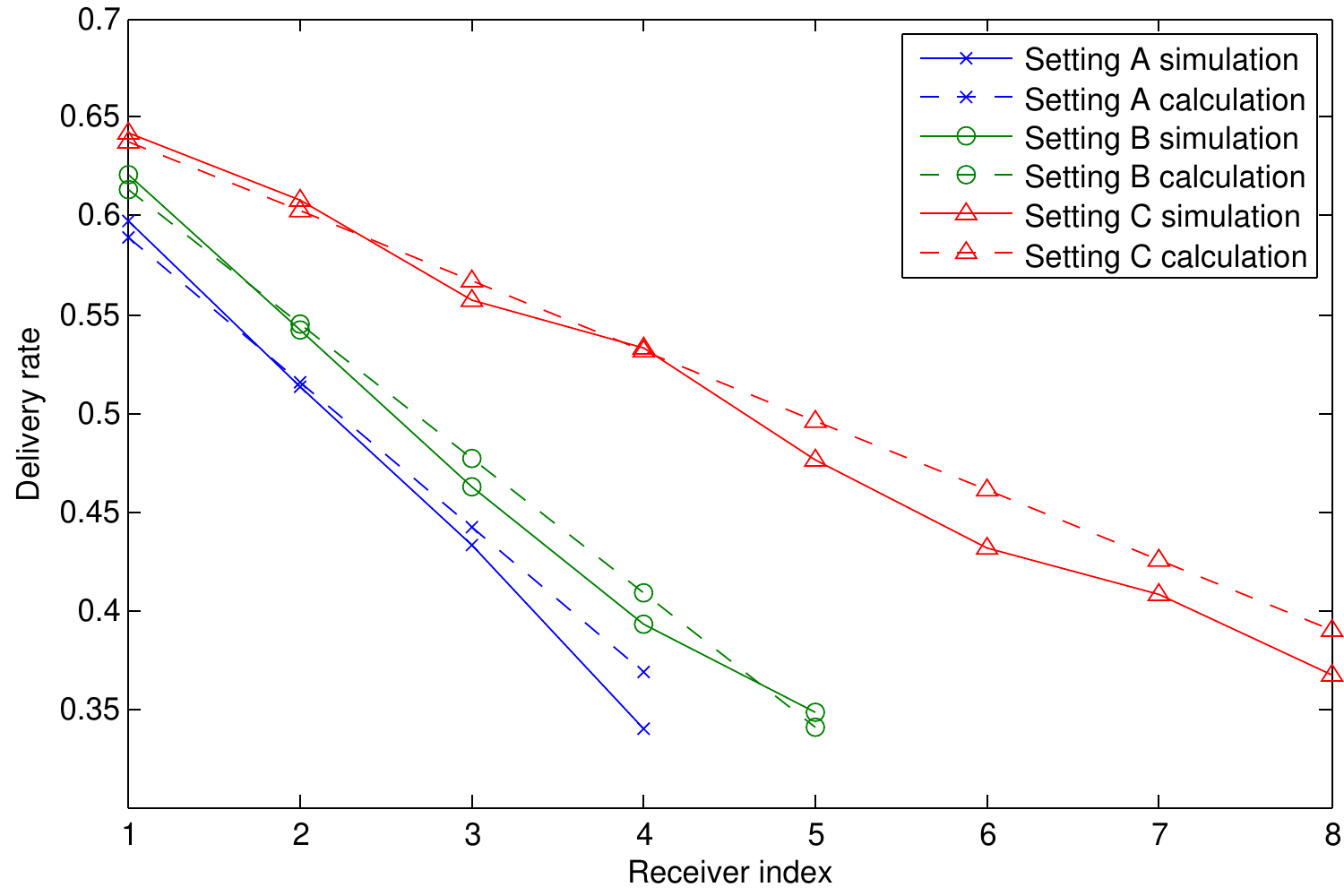}
\caption{Calculated and simulated delivery rates under the settings of Table \ref{tab:simsettings}, using the $\betav$ values derived in Algorithm \ref{alg:fairness}.}\label{fig:fairrates}
\end{center}\end{figure}

In Fig. \ref{fig:fairrates} we compare the calculated and simulated delivery rates under the values of $\betav$ determined from Algorithm \ref{alg:fairness}. Similar to Fig. \ref{fig:rates}, we observe that for low receiver indices, estimates match well with calculated values, while some loss of accuracy is observed for higher receiver indices.

%% file: kdiff_conclusion.tex
\section{Conclusion}

In this paper we have demonstrated that, by relaxing the complete information acquisition requirement, it is possible to exceed the min-cut capacity of a broadcast packet erasure network. By implementing a deterministic coding scheme, we were able to achieve multi-rate delivery of the same set of ordered information to a set of heterogeneous receivers, allowing the sender's transmission rate to exceed the channel rate of a number of receivers.

Despite the complexity associated with the analysis of deterministic coding schemes, through the use of some approximations we were able to obtain tractable estimates of the receivers' delivery rates. These estimates were shown to be reasonably accurate, although some loss of accuracy was observed with a larger number of receivers. Based on our \med{approximate} analysis, we introduced a fairness algorithm that determined how the sender should allocate its resources to achieve fair delivery rate performance for the receivers.

While the rate analysis we have presented in this paper lends insight into the process of multi-rate delivery, it is specific to the network coding scheme we have used. It remains unclear whether higher delivery rates could in fact be achieved using other deterministic transmission schemes. Future work should include the characterisation of the capacity of a broadcast network, given the relaxation of the complete information acquisition requirement.